\documentclass{article}
\usepackage{microtype}
\usepackage{graphicx}
\usepackage{subfigure}
\usepackage{booktabs} % for professional tables

\usepackage{hyperref}
\usepackage[accepted]{icml2024}
\usepackage[group-separator={,}]{siunitx}
\usepackage{booktabs} % For formal tables
\usepackage{color}
\usepackage{listings}
\usepackage{enumitem}
\usepackage{threeparttable}
\usepackage{multirow}% http://ctan.org/pkg/multirow
\usepackage{hhline}% http://ctan.org/pkg/hhline
\usepackage{subfigure}
\usepackage{array}
\usepackage{pifont}
\usepackage{booktabs}
\usepackage{dsfont}
\usepackage{microtype}
\usepackage{graphicx}
\usepackage{stmaryrd}
\usepackage{amsmath}

\usepackage{amssymb}
\usepackage{amsthm}
\usepackage{url}
\usepackage{bbm}
\usepackage{xspace}
\usepackage[normalem]{ulem}
\usepackage{booktabs, siunitx}
\usepackage{xcolor}
\usepackage{colortbl}
\usepackage{tkz-euclide}
\usepackage[framemethod=tikz]{mdframed}
\usepackage{empheq}
\usepackage[most]{tcolorbox}
\usepackage{adjustbox}
\usepackage{wrapfig}
\usepackage{pgfplots}
\usepgfplotslibrary{groupplots}
\usepackage{pgfplotstable}

\usepackage{balance}
\expandafter\def\expandafter\UrlBreaks\expandafter{\UrlBreaks%  save the current one
  \do\a\do\b\do\c\do\d\do\e\do\f\do\g\do\h\do\i\do\j%
  \do\k\do\l\do\m\do\n\do\o\do\p\do\q\do\r\do\s\do\t%
  \do\u\do\v\do\w\do\x\do\y\do\z\do\A\do\B\do\C\do\D%
  \do\E\do\F\do\G\do\H\do\I\do\J\do\K\do\L\do\M\do\N%
  \do\O\do\P\do\Q\do\R\do\S\do\T\do\U\do\V\do\W\do\X%
  \do\Y\do\Z}

\newcommand{\yhmodify}[2]{#2}

\expandafter\def\expandafter\UrlBreaks\expandafter{\UrlBreaks%  save the current one
  \do\a\do\b\do\c\do\d\do\e\do\f\do\g\do\h\do\i\do\j%
  \do\k\do\l\do\m\do\n\do\o\do\p\do\q\do\r\do\s\do\t%
  \do\u\do\v\do\w\do\x\do\y\do\z\do\A\do\B\do\C\do\D%
  \do\E\do\F\do\G\do\H\do\I\do\J\do\K\do\L\do\M\do\N%
  \do\O\do\P\do\Q\do\R\do\S\do\T\do\U\do\V\do\W\do\X%
  \do\Y\do\Z}

\definecolor{codegreen}{rgb}{0,0.6,0}
\definecolor{codegray}{rgb}{0.5,0.5,0.5}
\definecolor{codepurple}{rgb}{0.58,0,0.82}
\definecolor{backcolour}{rgb}{1,1,1}
 
\lstdefinestyle{mystyle}{
    backgroundcolor=\color{backcolour},   
    commentstyle=\color{codegreen},
    keywordstyle=\color{magenta},
    numberstyle=\tiny\color{codegray},
    stringstyle=\color{codepurple},
    basicstyle=\ttfamily\footnotesize,
    breakatwhitespace=false,         
    breaklines=true,                 
    captionpos=b,                    
    keepspaces=true,                 
    numbers=left,                    
    numbersep=5pt,                  
    showspaces=false,                
    showstringspaces=false,
    showtabs=false,           
    tabsize=2,
    xleftmargin=1.5em
}

\newtheorem{theorem}{Theorem}[section]

\newtheorem{example}{Example}[section]
\newtheorem{remark}{Remark}[section]

\newcommand{\largenumber}[1]{\num[group-separator={,}]{#1}}
\newcommand{\technique}{PECAN\xspace}
\newcommand{\perturb}{\pi}
\newcommand{\perturbtrain}{S_r}

\newcommand{\bfx}{\mathbf{x}}

\newcommand{\poisoned}[1]{\widetilde{#1}}
\newcommand{\poisonfeat}{s}
\newcommand{\poisonins}{r}

\newcommand{\secondy}{y'}
\newcommand{\firsty}{y^*}

\newcommand{\featflip}{\textsc{f}_{\poisonfeat}}
\newcommand{\featflipone}{\textsc{f}_{1}}
\newcommand{\featfliptwo}{\textsc{f}_{2}}
\newcommand{\featflipthree}{\textsc{f}_{3}}

\newcommand{\bfxtest}{\bfx}

\newcommand{\alg}{A}

\newcommand{\salg}{\bar{\alg}}

\newcommand{\model}[1]{A_{#1}}

\newcommand{\modelpred}[2]{\alg_{#1}(#2)}
\newcommand{\modelpredcert}[2]{\alg^{\perturb}_{#1}(#2)}

\newcommand{\smodelpred}[2]{\salg_{#1}(#2)}
\newcommand{\poisonset}{\perturbtrain(D)}

\newcommand{\poisonsetpoisoned}{\perturbtrain(\poisoned{D})}

\newcommand{\assumeradii}{R}

\newcommand{\dbackdoor}{D_\mathrm{bd}}
\newcommand{\dabstain}{D_\mathrm{abs}}
\newcommand{\dsafe}{D_\mathrm{safe}}
\newcommand{\Nfirsty}{N_1}
\newcommand{\Nsecondy}{N_2}
\newcommand{\Nabs}{N_3}
\newcommand{\poisonedNfirsty}{\poisoned{N}_{\firsty}}
\newcommand{\poisonedNsecondy}{\poisoned{N}_{\secondy}}
\newcommand{\certlabel}{\mathrm{cert}}
\newcommand{\abslabel}{\mathrm{abstain}}

\newcommand{\specialcase}{\diamond}
\newcommand{\newdthree}{\poisoned{D}_{3}^*}

\usepackage{amsmath}
\usepackage{amssymb}
\usepackage{mathtools}
\usepackage{amsthm}

\usepackage[capitalize,noabbrev]{cleveref}

\theoremstyle{plain}
\theoremstyle{definition}

\theoremstyle{remark}

\usepackage[textsize=tiny]{todonotes}

\icmltitlerunning{\technique: A Deterministic Certified Defense Against Backdoor Attacks}

\begin{document}

\twocolumn[
\icmltitle{\technique: A Deterministic Certified Defense Against Backdoor Attacks}

% It is OKAY to include author information, even for blind
% submissions: the style file will automatically remove it for you
% unless you've provided the [accepted] option to the icml2023
% package.

% List of affiliations: The first argument should be a (short)
% identifier you will use later to specify author affiliations
% Academic affiliations should list Department, University, City, Region, Country
% Industry affiliations should list Company, City, Region, Country

% You can specify symbols, otherwise they are numbered in order.
% Ideally, you should not use this facility. Affiliations will be numbered
% in order of appearance and this is the preferred way.
\icmlsetsymbol{equal}{*}

\begin{icmlauthorlist}
\icmlauthor{Yuhao Zhang}{uwm}
\icmlauthor{Aws Albarghouthi}{uwm}
\icmlauthor{Loris D'Antoni}{uwm}
\end{icmlauthorlist}

\icmlaffiliation{uwm}{University of Wisconsin-Madison}

\icmlcorrespondingauthor{Yuhao Zhang}{yuhaoz@cs.wisc.edu}

% You may provide any keywords that you
% find helpful for describing your paper; these are used to populate
% the "keywords" metadata in the PDF but will not be shown in the document
% \icmlkeywords{Machine Learning, ICML}

\vskip 0.3in
]

% this must go after the closing bracket ] following \twocolumn[ ...

% This command actually creates the footnote in the first column
% listing the affiliations and the copyright notice.
% The command takes one argument, which is text to display at the start of the footnote.
% The \icmlEqualContribution command is standard text for equal contribution.
% Remove it (just {}) if you do not need this facility.

\printAffiliationsAndNotice{}  % leave blank if no need to mention equal contribution
% \printAffiliationsAndNotice{\icmlEqualContribution} % otherwise use the standard text.

\begin{abstract}
Neural networks are vulnerable to backdoor poisoning attacks, where the attackers maliciously \emph{poison} the training set and insert \emph{triggers} into the test input to change the prediction of the victim model. 
% Existing certified defenses against backdoor attacks are probabilistic and ineffective, thus raising the question of whether one can design a deterministic certified approach to defend against backdoor attacks.
Existing defenses for backdoor attacks either provide no formal guarantees or come with expensive-to-compute and ineffective probabilistic guarantees.
We present \technique, an efficient and certified approach for defending against backdoor attacks.
The key insight powering \technique is to apply off-the-shelf test-time evasion certification techniques on a set of neural networks trained on disjoint \emph{partitions} of the data. 
We evaluate \technique on image classification and malware detection datasets. 
Our results demonstrate that \technique can (1) significantly outperform the state-of-the-art certified backdoor defense, both in defense strength and efficiency, and (2) on real backdoor attacks, \technique can reduce attack success rate by order of magnitude when compared to a range of baselines from the literature. 
\end{abstract}

\section{Introduction}
Deep learning models are vulnerable to \emph{backdoor} poisoning attacks~\cite{backdoor_hidden, backdoor_labelconsistent}, where the attackers can \emph{poison} a small fragment of the training set before model training and add \emph{triggers} to inputs at test time.
As a result, the prediction of the victim model that was trained on the poisoned training set will diverge in the presence of a trigger in the test input.

Effective backdoor attacks have been proposed for image recognition~\cite{badnets}, sentiment analysis~\cite{backdoor_NLP}, and malware detection~\cite{backdoor_Malwarepoison}. 
For example, the explanation-guided backdoor attack~(XBA)~\cite{backdoor_Malwarepoison} can break malware detection models as follows:
The attacker \emph{poisons} a small portion of benign software in the training set by modifying the values of the most important features so that the victim model recognizes these values as evidence of the benign prediction. 
At  test time, the attacker inserts a \emph{trigger} by changing the corresponding features of malware to camouflage it as benign software and making it bypass the examination of the victim model.
Thus, backdoor attacks are of great concern to the security of deep learning models and systems that are trained on data gathered from different sources, e.g., via web scraping.

We identify two limitations of existing defenses to backdoor attacks.
First, many existing approaches only provide empirical defenses that are specific to certain attacks and do not generalize to \emph{all} backdoor attacks. 
Second, existing certified defenses---i.e., approaches that produce robustness certificates---are either unable to handle backdoor attacks, or are probabilistic (instead of deterministic), and therefore expensive and ineffective in practice. 

\textbf{Why certification?}
A defense against backdoor attacks should construct effective \emph{certificates} (proofs) that the learned model can indeed defend against backdoor attacks.
Empirical defenses~\cite{heu_advtrain, heu_finetune} do not provide certificates, can only defend against specific attacks, and can be bypassed by new unaccounted-for attacks~\cite{break_backdoorfl, break_sanitization}.
In Section~\ref{sec: rq2}, we show that existing empirical defenses cannot defend against XBA when only 0.1\% of training data is poisoned.
Certification has been successful at building models that are provably robust to \emph{trigger-less} poisoning attacks and \emph{evasion} attacks, but models trained to withstand such attacks are still weak against backdoor attacks.
The \emph{trigger-less} attack~\cite{triggerless_convpolytope, triggerless_featurecollision} assumes the attacker can poison the training set but cannot modify the test inputs, e.g., adding triggers, while the \emph{evasion} attack~\cite{pgd, advpatch} assumes the attacker modifies the test inputs but cannot poison the training set.
\yhmodify{
Existing certified defenses against trigger-less and evasion attacks, e.g.,  DPA~\cite{partitioning} and CROWN-IBP~\cite{crownibp} ,
}{
Existing certified defenses against trigger-less attacks, e.g., DPA~\cite{partitioning}, and against evasion attacks, e.g., CROWN-IBP~\cite{crownibp} and PatchGuard++~\cite{patchguard++},
}
cannot defend against backdoor attacks as they can either defend against the poison in the training data or the triggers at test time, but not both.
As we show in the experiments, we can break these certified defenses using 
\yhmodify{}{BadNets~\cite{badnets} and} XBA (Section~\ref{sec: rq2}). 

\textbf{Why determinism?}
It is desirable for a certified defense to be \emph{deterministic} because probabilistic defenses~\cite{bagflip, RAB} typically require one to retrain thousands of models when performing predictions for a single test input.
Retraining can be mitigated by Bonferroni correction, which allows reusing the trained models for a fixed number of predictions.
However, retraining is still necessary after a short period, making it hard to deploy these defenses in practice. 
On the other hand, deterministic defenses~\cite{partitioning, dpa_improve} can reuse the trained models an arbitrary number of times when producing certificates for different test inputs.
Furthermore, probabilistic defenses for backdoor attacks, e.g., BagFlip~\cite{bagflip}, need to add noise to the training data, resulting in low accuracy for datasets that cannot tolerate too much noise when training (Section~\ref{sec: rq1}).
% Furthermore, the tighter confidence level enforced by the Bonferroni correction may harm the certified accuracy.

\begin{figure}
  \centering
  \input{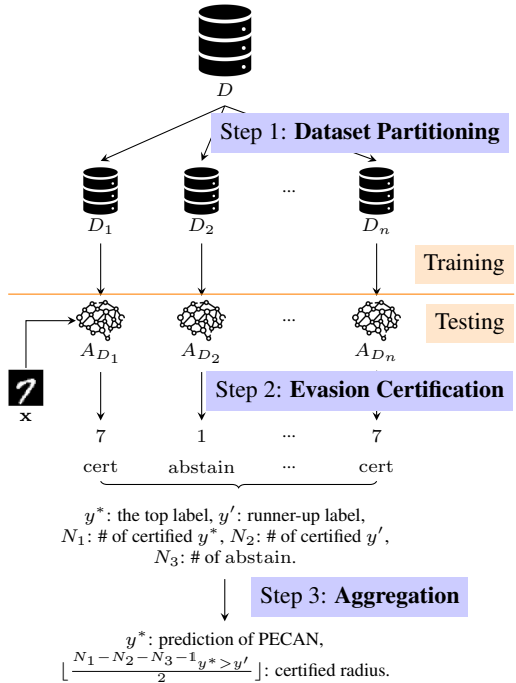}
  \caption{An overview of our approach \technique. 
  }
  \label{fig:overview}
  \vspace{-1em}
\end{figure}

\paragraph{\technique}
We present \technique (\textbf{P}artitioning data and \textbf{E}nsembling of \textbf{C}ertified neur\textbf{A}l \textbf{N}etworks), 
a deterministic certified defense against backdoor attacks for neural networks.
\technique can take \emph{any} off-the-shelf technique for evasion certification and use it to construct a certified backdoor defense. This insight results in a simple modular implementation that can leverage future advances in evasion certification algorithms.
Specifically, \technique trains a set of neural networks on disjoint partitions of the dataset, and then applies evasion certification to the neural networks. By partitioning the dataset, we analytically bound the number of poisoned data seen per neural network; by employing evasion certification, we bound the number of neural networks that are robust in the face of triggers.
Using this information, we efficiently derive a backdoor-robustness guarantee.

Figure~\ref{fig:overview} illustrates the workflow of \technique.
In Step 1, inspired by \emph{deep partition aggregation}~\cite{partitioning}, \technique deterministically partitions a dataset into multiple disjoint subsets. This step ensures that a poisoned data item only affects a single partition.
In Step 2, \technique trains an ensemble of neural networks, one on each partition.
At test time, \technique performs evasion certification to check which neural networks are immune to triggers; those that are not immune (or that cannot be proven immune) abstain from performing a prediction.
Finally, in Step 3, \technique aggregates the results of the ensemble and produces a prediction together with a robustness certificate: the percentage of the poisoned data in the training set that the training process can tolerate, the \emph{certified radius}.

We evaluate \technique on three datasets, MNIST, CIFAR10, and EMBER. 
\technique outperforms or competes with BagFlip, the state-of-the-art probabilistic certified defense against backdoor attacks. 
Furthermore, BagFlip takes hours to compute the certificate, while \technique takes seconds.
\yhmodify{
Second, when we evaluate \technique against a known backdoor attack XBA~\cite{backdoor_Malwarepoison},
\technique reduces the attack success rate from 66.37\% to 1.18\%, while other approaches fail to defend against the backdoor attack.
}
{
We also evaluate \technique against two known backdoor attacks, BadNets~\cite{badnets} and XBA~\cite{backdoor_Malwarepoison},
\technique reduces the average attack success rates from 90.24\% to 0.67\% and 66.37\% to 2.19\%, respectively.
}
\section{Related Work}
Deep learning models are vulnerable to backdoor attacks~\cite{backdoor_hidden, backdoor_labelconsistent}, and empirical defenses~\cite{heu_advtrain, heu_finetune} can be bypassed~\cite{break_backdoorfl, break_sanitization}. Hence, our focus on building a certified defense. 

\paragraph{Certified defenses against backdoor attacks}
Existing certification approaches provide probabilistic certificates by extending randomized smoothing~\cite{randomizesmoothing, divergences, tightflipbound}, originally proposed to defend against adversarial evasion attacks, to defend against backdoor attacks.
BagFlip~\cite{bagflip} is the state-of-the-art model-agnostic probabilistic defense against feature-flipping backdoor attacks. 
\citet{featureflip, RAB} proposed backdoor-attack defenses that are also model-agnostic, but are less effective than BagFlip.
\technique is deterministic and therefore less expensive and more effective than these defenses.
Probabilistic defenses are model-agnostic; while \technique is evaluated on neural networks, it can work for any machine learning model as that supports a deterministic evasion certification approach. 
% We regard extending \technique to other machine learning models as future work.
\citet{certkrNN} proposed a deterministic de-randomized smoothing approach for kNN classifiers.
Their approach computes the certificates using an expensive dynamic programming algorithm, whereas \technique's certification algorithm has constant time complexity.
XRand~\cite{xrand} presents a certified defense against XBA by leveraging differential privacy. 
Unlike \technique, XRand adopts a different attack model, and its certifications are probabilistic in nature.
FPA~\cite{fpa} employs a deterministic certified defense mechanism that partitions the feature space instead of the dataset. 
% However, FPA does not provide protection against data poisoning attacks that involve modifications to training labels or the insertion/removal of training examples.
We discuss the limitation of FPA and compare it with \technique in Section~\ref{sec: rq2}.

% \paragraph{Certified backdoor attack detection}
% CBD~\citep{cbd} and PECAN are two methods that address backdoor attacks in machine learning, but they have different goals, assumptions, and application scenarios. CBD is a certified backdoor detector that aims to identify whether a trained classifier is backdoored, assuming the defender has access to a small, clean validation set. On the other hand, PECAN is a certified defense that trains a classifier with a potentially poisoned dataset while providing certifications for each prediction, guaranteeing robustness against backdoor attacks up to a certain poisoning rate. PECAN assumes the defender has access to the dataset and the training process. Due to the difference in the difficulty of their tasks, CBD demonstrates higher performance in terms of certified true positive rate compared to PECAN's certified accuracy.

\paragraph{Certified defenses against trigger-less attacks}
Many approaches provide certificates for trigger-less attacks.
\citet{bagging} use bootstrap aggregating (Bagging).
\citet{framework_def} extended Bagging with new selection strategies.
\citet{labelflip} defend against label-flipping attacks on linear classifiers.
Differential privacy~\cite{differential} can also provide probabilistic certificates for trigger-less attacks.
% The other line of work provides deterministic certificates for trigger-less attacks. 
DPA~\cite{partitioning} is a deterministic defense that partitions the training set and ensembles the trained classifiers. 
\citet{dpa_improve} proposed FA, an extension of DPA, by introducing a spread stage.
A conjecture proposed by \citet{conjecture} implies that DPA and FA are asymptotically optimal defenses against trigger-less attacks. 
\citet{collective} proposed to compute collective certificates, while \technique computes sample-wise certificates. 
\citet{certkrNN, prog_decisiontree, decisiontree} provide certificates for nearest neighborhood classifiers and decision trees.
The approaches listed above only defend against \textit{trigger-less} attacks, while \technique is a deterministic approach for \textit{backdoor} attacks.

\paragraph{Certified backdoor attack detection}
CBD~\citep{cbd} and \technique are two methods that address backdoor attacks, but they have different goals, assumptions, and application scenarios. 
CBD is a certified backdoor detector that aims to identify whether a trained classifier is backdoored, assuming the defender has access to a small, clean validation set. 
On the other hand, \technique is a certified defense that trains a classifier with a potentially poisoned dataset while providing certifications for each prediction, guaranteeing robustness against backdoor attacks up to a certain poisoning rate. 
\technique assumes the defender has access to the dataset and the training process. 
Certified detection is an easier task than certified classification\citep{easier}. Intuitively, certified detection only needs to determine if the model is backdoored, while certified classification requires training a classifier with a poisoned dataset. 
Due to the difference in the difficulty of their tasks, CBD demonstrates higher performance in terms of certified true positive rate compared to \technique's certified accuracy.

\paragraph{Certified defenses against evasion attacks}
There are two lines of certified defense against evasion attacks: complete certification~\cite{cuttingplane, Marabou} and incomplete certification~\cite{certlstm, deeppoly}. 
The complete certified defenses either find an adversarial example or generate proof that all inputs in the given perturbation space will be correctly classified.
Compared to the complete certified defenses, the incomplete ones will abstain from predicting if they cannot prove the correctness of the prediction because their techniques will introduce over-approximation.
% The complete approaches do not have over-approximation issues but require expensive verification algorithms such as branch and bound. 
Our implementation of PECAN uses an incomplete certified approach CROWN-IBP~\cite{crownibp} because it is the best incomplete approach, trading off between efficiency and the degree of over-approximation.
\section{Problem Definition}
\label{sec: def}
% \yh{define the algorithm please check.}
Given a dataset $D = \{(\bfx_1, y_1), \ldots, (\bfx_n, y_n)\}$, a (test) input $\bfxtest$, and a machine learning algorithm $\alg$, we write $\model{D}$ to denote the machine learning model learned on dataset $D$ by the algorithm $\alg$, and $\modelpred{D}{\bfxtest}$ to denote the output label predicted by the model $\model{D}$ on input $\bfxtest$. 
We assume the algorithm will behave the same if trained on the same dataset across multiple runs. 
This assumption can be guaranteed by fixing the random seeds during training. 

We are interested in certifying that if an attacker has poisoned the dataset, the model we have trained on the dataset will still behave ``well'' on the test input with maliciously added triggers.
Before describing what ``well'' means, we need to define the \emph{perturbation spaces} of the dataset and the test input, i.e.,  what possible changes the attacker could make to the dataset and the test input. 

\paragraph{Perturbation space of the dataset}
Following \citet{partitioning}, we define a \emph{general} perturbation space over the dataset, allowing attackers to delete, insert, or modify training examples in the dataset.
Given a dataset $D$ and a \textit{radius} $\poisonins\geq 0$, we define the \textit{perturbation space} as the set of datasets that can be obtained by deleting or inserting up to $\poisonins$ examples in $D$:
\[
\poisonset = \left\{\poisoned{D} \mid |D \ominus \poisoned{D}|\le \poisonins\right\},
\]
where $A \ominus B$ is the symmetric difference of sets $A$ and $B$.
Intuitively, $\poisonins$ quantifies how many examples need to be deleted or inserted to transform from $D$ to $\poisoned{D}$.

\begin{example}
\label{exp: poisonset_approach}
    If the attacker modifies one training example $\bfx \in D$ to another training example $\poisoned{\bfx}$ to form a poisoned dataset $\poisoned{D} = (D \setminus \{\bfx\}) \cup \{\poisoned{\bfx}\}$.
    Then $\poisoned{D} \in S_2(D)$ but $\poisoned{D} \notin S_1(D)$ because $\poisonset$ considers one modification as one deletion and one insertion.
\end{example}

Note that we assume a more general perturbation space of the training set than the ones considered by BagFlip~\cite{bagflip} and FPA~\cite{fpa}; our work allows inserting and deleting examples instead of just modifying existing training examples.

\paragraph{Perturbation space of the test input}
We write $\perturb(\bfxtest)$ to denote the set of perturbed examples that an attacker can transform the example $\bfxtest$ into.
Formally, the perturbation space $\perturb(\bfxtest)$ can be defined as the $l_p$ norm ball with radius $\poisonfeat$ around the test input $\bfxtest$,
\[\perturb(\bfx) = \{\poisoned{\bfx}\mid \|\bfx - \poisoned{\bfx}\|_p \le \poisonfeat\}\]

\begin{example}
\label{exp: l0_attack}
    An instantiation of $\perturb(\bfxtest)$ is the $l_0$ feature-flip perturbation $\featflip(\bfx)$, which allows the attacker to modify up to $\poisonfeat$ features in an input $\bfx$,
    \[\featflip(\bfx) = \{\poisoned{\bfx}\mid \|\bfx - \poisoned{\bfx}\|_0 \le \poisonfeat\}\]
\end{example}

Dynamic backdoor attacks~\cite{dynamic_trigger}
\yhmodify{}{like BadNets~\cite{badnets}}, which involve placing different patches in an input $\bfx$, with each patch having a maximum size of $\poisonfeat$, can be captured by $\featflip(\bfx)$.
XBA~\cite{backdoor_Malwarepoison}, which modifies a predetermined set of features up to $\poisonfeat$ features in an input $\bfx$, can also be captured by $\featflip(\bfx)$. 
Note that we assume a more general perturbation space of the test input than the one considered by FPA, which cannot handle dynamic backdoor attacks.

\paragraph{Threat models}
% Next, we define what type of guarantees we are interested in our learning algorithm and model. 
% 
We consider backdoor attacks, where the attacker can perturb both the training set and the test input, but cannot control the training process of models.
For the training set, we assume the attacker selects a poisoned training set from a perturbation space $\poisonset$ of the training set $D$ with a radius $\poisonins\geq 0$.
For the test input, we assume the attacker selects a test input with a malicious trigger from a perturbation space $\perturb(\bfx)$ of the test input $\bfx$ with a given $l_p$ norm and the radius $\poisonfeat$.

We say that an algorithm $\alg$ is robust to a \textbf{backdoor attack} on a backdoored test input $\poisoned{\bfxtest}$ if the algorithm trained on any perturbed dataset $\poisoned{D}$ would predict the backdoored input $\poisoned{\bfxtest}$ the same as $\modelpred{D}{\bfxtest}$. 
Formally,
\begin{align}
     \forall \poisoned{D} \in \poisonset,\ \poisoned{\bfxtest} \in \perturb(\bfxtest).\ \modelpred{\poisoned{D}}{\poisoned{\bfxtest}} = \modelpred{D}{\bfxtest}\label{eq:goal}
\end{align}
% \yh{add the above definition of $\mathrm{proj}_{\bfx}$.}

\begin{remark}
    When $\poisonins = 0$, Eq~\ref{eq:goal} degenerates to evasion robustness, i.e., $\forall \poisoned{\bfxtest} \in \perturb(\bfxtest).\ \modelpred{D}{\poisoned{\bfxtest}} = \modelpred{D}{\bfxtest}$, because $S_0(D) = \{D\}$.
\end{remark}

Given a large enough radius $r$, an attacker can always change enough inputs and succeed at breaking robustness. 
Therefore, we will typically focus on computing the maximal radius $r$ for which we can prove that Eq~\ref{eq:goal} for given perturbation spaces $\poisonset$ and $\perturb(\bfxtest)$.
We refer to this quantity as the \textit{certified radius}.

\paragraph{Certified guarantees}
% \loris{earlier we didn't say that A was an algorithm, might be worth stating that at the beginning of sec 3}
This paper aims to design a certifiable algorithm $\alg$, which can defend against backdoor attacks, and to compute the certified radius of $\alg$.
In our experiments (Section~\ref{sec: rq1}), we suppose a given benign dataset $D$ and a benign test input $\bfx$, and we certifiably quantify the robustness of the algorithm $\alg$ against backdoor attacks by computing the certified radius. 

In Section~\ref{sec: rq2}, we also experiment with how the certifiable algorithm $\alg$ defends backdoor attacks if a poisoned dataset $\poisoned{D}$ and a test input $\poisoned{\bfx}$ with malicious triggers are given, but the clean data is unknown.
We theoretically show that we can still compute the certified radius if the clean data $D$ and $\bfx$ are unknown in Section~\ref{sec: inversed_problem}.
\section{The \technique Certification Technique}
\label{sec: approach}

Our approach, which we call \technique (\textbf{P}artitioning data and \textbf{E}nsembling of \textbf{C}ertified neur\textbf{A}l \textbf{N}etworks), is a deterministic certification technique that defends against backdoor attacks.
Given a learning algorithm $\alg$, we show how to automatically construct a new learning algorithm $\salg$ with certified backdoor-robustness guarantees (Eq~\ref{eq:goal}) in Section~\ref{sec: construct}.
% In Section~\ref{sec: proof}, we prove the certified backdoor-robustness guarantees (Eq~\ref{eq:goal}) provided by $\salg$.
We further discuss how $\salg$ can defend against a backdoored dataset and formally justify our discussion in Section~\ref{sec: inversed_problem}.

\subsection{Constructing Certifiable Algorithm $\salg$}
\label{sec: construct}

% \yh{added the intuition.}
The key idea of \technique is that we can take any off-the-shelf technique for evasion certification and use it to construct a certified backdoor defense.
Intuitively, \technique uses the evasion certification to defend against the possible triggers at test time, and it encapsulates the evasion certification in deep partition aggregation (DPA)~\cite{partitioning} to defend against training set poisoning.

Given a dataset $D$, a test input $\bfxtest$, and a machine learning algorithm $\alg$, \technique produce a new learning algorithm $\salg$ as described in the following steps (shown in Figure~\ref{fig:overview}),

    \paragraph{Dataset Partitioning}
    We partition the dataset $D$ into $n$ disjoint sub-datasets, denoted as $D_1, \ldots, D_n$, using a hash function that deterministically maps each training example into a sub-dataset $D_i$.
    Train $n$ classifiers $\model{D_1}, \ldots, \model{D_n}$ on these sub-datasets. 
    
    \paragraph{Evasion Certification} We certify whether the prediction of each classifier $\model{D_i}$ is robust under the perturbation space $\perturb(\bfx)$ by any evasion certification approach for the learning algorithm, e.g., CROWN-IBP for neural networks~\cite{lirpa}. 
    Formally, the certification approach determines whether the following equation holds, 
    \begin{align}
        \forall \poisoned{\bfxtest} \in \perturb(\bfx).\ \modelpred{D_i}{\bfxtest} = \modelpred{D_i}{\poisoned{\bfxtest}} \label{eq: ibp_goal}
    \end{align}
    We denote the output of each certification as $\modelpredcert{D_i}{\bfxtest}$, which can either be $\modelpredcert{D_i}{\bfxtest} = \certlabel$, meaning Eq~\ref{eq: ibp_goal} is certified. 
    Otherwise, $\modelpredcert{D_i}{\bfxtest}=\abslabel$, meaning the certification approach cannot certify Eq~\ref{eq: ibp_goal}.

    \paragraph{Aggregation}
    We compute the top label $\firsty$ by aggregating all predictions from 
    $\modelpred{D_i}{\bfxtest}$.
    % \loris{shouldn't you only count ones that didn't abstain both for y and y*?}
    Concretely, $\firsty \triangleq \underset{y \in \mathcal{C}}{\mathrm{argmax}} \sum_{i=1}^n \mathds{1}_{\modelpred{D_i}{\bfxtest} = y},$
    where $\mathcal{C}=\{0,1,\ldots\}$ is the set of possible labels.
    Note that if a tie happens when taking the argmax, we break ties deterministically by setting the smaller label index as $\firsty$.
    We denote the runner-up label as $\secondy$ as 
    % \loris{should be y' instead of y at end of Eq?}
    $\underset{y \in \mathcal{C} \wedge y \neq \firsty}{\mathrm{argmax}} \sum_{i=1}^n \mathds{1}_{\modelpred{D_i}{\bfxtest} = y}.$
    % Note that if a tie happens when taking the argmax, we break ties deterministically by setting the smaller label index as $\secondy$.
    We count the number of certified predictions equal to $\firsty$ as $\Nfirsty$, the number of certified predictions equal to $\secondy$ as $\Nsecondy$, and the number of abstentions as $\Nabs$.

    We set the prediction $\smodelpred{D}{\bfxtest}$ as $\firsty$.
    We compute the certified radius $\poisonins$ in the following two cases. 
    If $\Nfirsty - \Nsecondy - \Nabs - \mathds{1}_{\firsty > \secondy} < 0$, we set $\poisonins$ as $\specialcase$, i.e., a value denoting no certification.
    In this case, \technique cannot certify that $\salg$ is robust to evasion attacks even if the dataset is not poisoned.
    Otherwise, we compute $\poisonins$ as $\lfloor \frac{\Nfirsty - \Nsecondy - \Nabs - \mathds{1}_{\firsty > \secondy}}{2}\rfloor$.
A special case is $\poisonins=0$, when \technique can certify $\salg$ is robust to evasion attacks, but cannot certify that it is robust if the dataset is poisoned.

We note that the computation of the certified radius is equivalent to DPA when no classifier abstains, i.e., $\Nabs=0$,

% \subsection{Proving the Soundness of \technique}
% \label{sec: proof}
% In this section, we show that the prediction $\smodelpred{D}{\bfxtest}$ and the certified radius $\poisonins$ satisfy the certified backdoor-robustness guarantees (Eq~\ref{eq:goal}) by proving the following theorem. 
\begin{theorem}[Soundness of \technique]
\label{theorem: main}
Given a dataset $D$ and a test input $\bfxtest$, \technique computes the prediction $\smodelpred{D}{\bfxtest}$ and the certified radius as $\poisonins$.
Then, either $\poisonins=\specialcase$ or
\begin{align}
     \forall \poisoned{D} \in \poisonset,\ \poisoned{\bfxtest} \in \perturb(\bfx).\ \smodelpred{\poisoned{D}}{\poisoned{\bfxtest}} = \smodelpred{D}{\bfxtest} \label{eq:dab_cert}
\end{align}
\end{theorem}

We provide a proof sketch of Theorem~\ref{theorem: main}, and the main proof can be found in Appendix~\ref{sec: appendix_main_proof}.
The key idea is that if we can prove that a majority of classifiers are immune to poisoned data, we can prove that the aggregated result is also immune to poisoned data. 
We start by lower bounding the number of classifiers 
% (the green part in Figure~\ref{fig:proof}) 
that are immune to poisoned data as $\Nfirsty - \poisonins$ and upper bounding the number of classifiers that can be manipulated by the attackers 
% (the gray and red parts in Figure~\ref{fig:proof}) 
as $\Nsecondy + \poisonins + \Nabs$.
Then, we show that if $\poisonins \le \lfloor \frac{\Nfirsty - \Nsecondy - \Nabs - \mathds{1}_{\firsty > \secondy}}{2}\rfloor$, the classifiers immune to poisoned data will always be a majority.

\subsection{\technique under the Backdoored Data}
\label{sec: inversed_problem}
The above algorithm and proof of \technique assume that a clean dataset $D$ and a clean test example $\bfxtest$ are already given. 
However, we may be interested in another scenario where the poisoned dataset $\poisoned{D} \in \poisonset$ and the input example $\poisoned{\bfxtest} \in \perturb(\bfxtest)$ with malicious triggers are given, and the clean data $D$ and $\bfxtest$ are unknown.
In other words, we want to find the maximal radius $\poisonins$ such that $\smodelpred{\poisoned{D}}{\poisoned{\bfxtest}} = \smodelpred{D}{\bfxtest}$ for any $D$ and $\bfxtest$ that can be perturbed to $\poisoned{D}$ and $\poisoned{\bfxtest}$ by the perturbation $\perturbtrain$ and $\perturb$, respectively.
Formally,
\begin{small}
\begin{align}
     \forall D, \bfxtest.\ \poisoned{D} \in \poisonset \wedge \poisoned{\bfx} \in \perturb(\bfxtest) \implies \smodelpred{\poisoned{D}}{\poisoned{\bfxtest}} = \smodelpred{D}{\bfxtest}\label{eq:inverse_goal}
\end{align}
\end{small}%

Intuitively, Eq~\ref{eq:inverse_goal} is the symmetrical version of Eq~\ref{eq:goal}.
Owing to the symmetrical definition of $\perturbtrain$ and $\perturb$, if we apply \technique to the given poisoned data $\poisoned{D}, \poisoned{\bfxtest}$, then the prediction $\smodelpred{\poisoned{D}}{\poisoned{\bfxtest}}$ and the certified radius $\poisonins$ satisfy the certified backdoor-robustness guarantee (Eq~\ref{eq:inverse_goal}).
The following theorem formally states the soundness of \technique under the backdoored data. We prove Theorem~\ref{theorem: inversed} in Appendix~\ref{sec: proof_inversed}.

\begin{theorem}[Soundness of \technique under backdoored data]
\label{theorem: inversed}
Given a dataset $\poisoned{D}$ and a test input $\poisoned{\bfxtest}$, \technique computes the prediction $\smodelpred{\poisoned{D}}{\poisoned{\bfxtest}}$ and the certified radius $\poisonins$.
Then, either $\poisonins=\specialcase$ or Eq~\ref{eq:inverse_goal} holds.
\end{theorem}

\section{Experiments}
We implemented \technique in Python and provided the implementation in the supplementary materials.

In Section~\ref{sec: rq1}, we evaluate the effectiveness and efficiency of \technique by comparing it to BagFlip~\cite{bagflip}, the state-of-the-art probabilistic certified defense against backdoor attacks. We use CROWN-IBP, implemented in auto-LiRPA~\cite{lirpa}, as the evasion defense approach in this setting. Whenever we use CROWN-IBP for the evasion defense approach, we also use it to train the classifiers in the dataset-partitioning step since the classifiers trained by CROWN-IBP can improve the certification rate in the evasion-certification step.

In Section~\ref{sec: rq2}, we evaluate the effectiveness of \technique under the patch attack using BadNets~\cite{badnets} for image classification and the explanation-guided backdoor attack (XBA)~\cite{backdoor_Malwarepoison} for malware detection and compare \technique to other approaches.
We use PatchGuard++~\cite{patchguard++} as the evasion defense approach for image classification and use CROWN-IBP as the evasion defense approach for malware detection.

\subsection{Experimental Setup}
\paragraph{Datasets}
We conduct experiments on MNIST, CIFAR10, CIFAR10-02, and  EMBER~\cite{ember} datasets.
CIFAR10-02~\cite{RAB} is a subset of CIFAR10, comprising examples labeled as 0 and 2. It consists of 10,000 training examples and 2,000 test examples.
EMBER is a malware detection dataset containing 600,000 training and 200,000 test examples. Each example is a vector containing 2,351 features of the software, e.g., number of sections and number of writeable sections.

\paragraph{Models} 
When comparing \technique and BagFlip, we train fully connected neural networks with four layers for MNIST and CIFAR10 datasets.
For experiments using PatchGuard++, we use the BagNet~\cite{bagnet} model structure used by PatchGuard++.
We use the same fully connected neural network for EMBER as in related works~\cite{bagflip, backdoor_Malwarepoison}.
We use the same data augmentation for \technique and other baselines.

\paragraph{Metrics}
For each test input $\bfxtest_i,y_i$, the \technique will predict a label and the certified radius $\poisonins_i$.
In this section, we assume that the attacker \emph{modified} $\assumeradii\%$ examples in the training set.
We denote $\assumeradii$ as the \emph{modification amount}.
We summarize all the used metrics as follows:
% \loris{is the use of the percentage sign correct int he formulas below? It feels awkward}

\noindent \textit{Certified Accuracy} denotes the percentage of test examples that are correctly classified and whose certified radii are no less than $\assumeradii$, i.e., $\frac{1}{m}\sum_{i=1}^m \mathds{1}_{\smodelpred{D}{\bfxtest_i} = y_i \wedge \frac{\poisonins_i}{|D|} \ge 2\assumeradii\%}$, where $m$ and $|D|$ are the sizes of test and training set, respectively.
Notice that there is a factor of $2$ on the modification amount $\assumeradii$ because $\poisonset$ considers one modification as one insertion and one deletion, as in Example~\ref{exp: poisonset_approach}.

\noindent  \textit{Normal Accuracy} denotes the percentage of test examples that are correctly classified by the algorithm \emph{without} certification, i.e., $\frac{1}{m}\sum_{i=1}^m \mathds{1}_{\smodelpred{D}{\bfxtest_i} = y_i}$.

\noindent  \textit{Attack Success Rate (ASR)}
We are interested in how many test examples are originally correctly classified without the malicious trigger but wrongly classified after adding the trigger, i.e., $\frac{1}{m}\sum_{i=1}^m \mathds{1}_{\smodelpred{D}{\poisoned{\bfxtest}_i} \neq y_i \wedge \smodelpred{D}{\bfxtest_i} = y_i \wedge \frac{\poisonins_i}{|D|} \ge 2\assumeradii\%}$, where $\bfxtest_i$ is the original test and $\poisoned{\bfxtest}_i$ is with a malicious trigger. 
% We denote the above quantity as the attack success rate.
% We note that a prediction can still be incorrect even if it is certified by \technique because the classifier can have incorrect predictions even when the data is clean.

\noindent  \textit{Abstention Rate} is computed as $\frac{1}{m}\sum_{i=1}^m \mathds{1}_{\frac{\poisonins_i}{|D|} < 2\assumeradii\%}$.
% \loris{is this correct? I'm not sure what it's counting}

\subsection{Effectiveness and Efficiency of \technique}
\label{sec: rq1}
We evaluate the effectiveness and efficiency of \technique on MNIST, CIFAR10, and EMBER under the backdoor attack with the $l_0$ feature-flip perturbation $\featflipone$, which allows the attacker to modify up to one feature in an example.
% \loris{argue why this is not a weak attack}
We compare \technique to BagFlip, the state-of-the-art probabilistic certified defense against $l_0$ feature-flip backdoor attacks. 
% Moreover, we note that \technique needs to construct harder proofs than BagFlip because their definitions of perturbation space are different, as discussed in Appendix~\ref{sec: appendix_detail_setup}.
In Appendix~\ref{sec: appendix_exp1}, we also evaluate the effectiveness of \technique against the perturbation space with the $l_\infty$ norm.

\begin{tcolorbox}[sharp corners, boxsep=0pt,left=3pt,right=3pt,top=2pt,bottom=2pt]
\paragraph{Summary of the results}
\technique achieves significantly higher certified accuracy than BagFlip on CIFAR10, EMBER, and MNIST.
\technique has similar normal accuracy as BagFlip for all datasets.
\technique is more efficient than BagFlip at computing the certified radius.
\end{tcolorbox}

\paragraph{Setup}
For \technique, we vary $n$, the number of partitions, to ensure a fair comparison with BagFlip. 
Appendix~\ref{sec: appendix_detail_setup} presents a detailed discussion of hyperparameter settings for BagFlip and \technique.
We denote \technique with different settings of $n$ as \technique-$n$.

BagFlip achieves meaningful results only on MNIST, where we tune the parameter $n$ of \technique to $3000$ to achieve the same normal accuracy of BagFlip and compare their results following the practice by~\citet{certkrNN,bagging}.

\paragraph{Results}

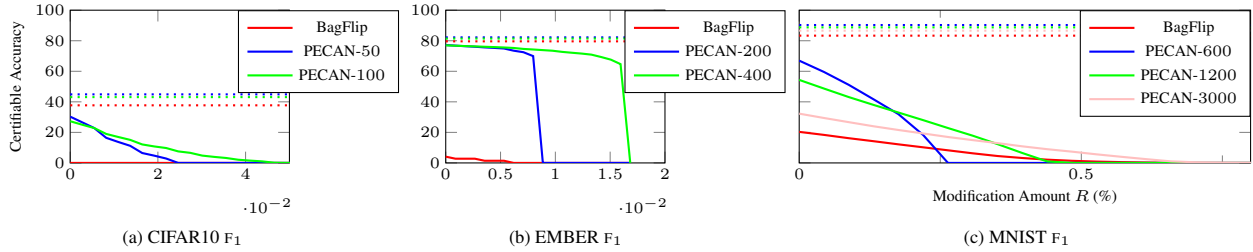
\begin{figure*}
\centering
\tiny
 \setlength{\tabcolsep}{1pt}
\begin{tabular}{ccc}
    \trimbox{0cm 0cm 0.3cm 0cm}{
\pgfplotsset{filter discard warning=false}
\pgfplotsset{every axis legend/.append style={
at={(1.5,1)},
anchor=north east}} 

\pgfplotscreateplotcyclelist{whatever}{%
    red,thick,every mark/.append style={fill=blue!80!black},mark=none\\%
    blue,thick,every mark/.append style={fill=red!80!black},mark=none\\%
    green,thick,every mark/.append style={fill=blue!80!black},mark=none\\%
    red,thick,dotted,every mark/.append style={fill=blue!80!black},mark=none\\%
    blue,thick,dotted,every mark/.append style={fill=red!80!black},mark=none\\%
    green,thick,dotted,every mark/.append style={fill=blue!80!black},mark=none\\%
    % gray,thick,dashed,every mark/.append style={fill=red!80!black},mark=none\\%
    % black,thick,every mark/.append style={fill=blue!80!black},mark=none\\%
    % gray,thick,every mark/.append style={fill=red!80!black},mark=none\\%
    % % brown!60!black,every mark/.append style={fill=brown!80!black},mark=triangle*\\%
    % black,mark=star\\%
    % blue,every mark/.append style={fill=blue!80!black},mark=diamond*\\%
    % red,densely dashed,every mark/.append style={solid,fill=red!80!black},mark=*\\%
    % brown!60!black,densely dashed,every mark/.append style={solid,fill=brown!80!black},mark=square*\\%
    % black,densely dashed,every mark/.append style={solid,fill=gray},mark=triangle*\\%
    % blue,densely dashed,mark=star,every mark/.append style=solid\\%
    % red,densely dashed,every mark/.append style={solid,fill=red!80!black},mark=diamond*\\%
    }
    
\begin{tikzpicture}
    \begin{groupplot}[
            group style={
                group size=1 by 1,
                horizontal sep=.05in,
                vertical sep=.05in,
                ylabels at=edge left,
                yticklabels at=edge left,
                xlabels at=edge bottom,
                xticklabels at=edge bottom,
            },
            height=.8in,
            xlabel near ticks,
            ylabel near ticks,
            scale only axis,
            width=0.17*\textwidth,
            xmin=0,
            ymin=0,
            ymax=100,
            %xmax=1.1
        ]

        \nextgroupplot[
            % title=Robustness by demographic group,
            ylabel=Certifiable Accuracy,
            xmax=0.05,
            % xtick={0,0.02,0.04},
            % minor xtick={0.01,0.03},
            cycle list name=whatever,
            xlabel=]
        \addplot table [x=x,y=2, col sep=comma]{data/cifar10.csv};
        \addplot table [x=x,y=0, col sep=comma]{data/cifar10.csv};
        \addplot table [x=x,y=1, col sep=comma]{data/cifar10.csv};
        \addplot table [x=x,y=2-normal, col sep=comma]{data/cifar10.csv};
        \addplot table [x=x,y=0-normal, col sep=comma]{data/cifar10.csv};
        \addplot table [x=x,y=1-normal, col sep=comma]{data/cifar10.csv};
        
        % \nextgroupplot[
        %     xmax=0.02,
        %     % xtick={0,0.005,0.01,0.015},
        %     % minor xtick={0.0025,0.0075,0.0125,0.0175},
        %      cycle list name=whatever,
        % ]
        % \addplot table [x=x,y=2, col sep=comma]{data/ember.csv};
        % \addplot table [x=x,y=0, col sep=comma]{data/ember.csv};
        % \addplot table [x=x,y=1, col sep=comma]{data/ember.csv};
        % \addplot table [x=x,y=2-normal, col sep=comma]{data/ember.csv};
        % \addplot table [x=x,y=0-normal, col sep=comma]{data/ember.csv};
        % \addplot table [x=x,y=1-normal, col sep=comma]{data/ember.csv};

\legend{BagFlip, \technique-$50$, \technique-$100$};

%  \node (P1) at (pvgroup c5r1.north east) {};
%     \node (P2) at (pvgroup c5r1.south east) {};
%     \path (P1) -- node[right]{\pgfplotslegendfromname{singlelegend}} (P2);
    
    \end{groupplot}
    \node at (1.5,-1) {\scriptsize (a) CIFAR10 $\featflipone$};
    % \node at (3.7,-1) {\scriptsize (b) EMBER $\featflipone$};
    % \node[draw] at (7.5,1.7) {\scriptsize (c) EMBER $\featflipone$};
    % \node[draw] at (9.8,0.5) {\scriptsize (d) EMBER $\featflipinf$};

\end{tikzpicture}
} & \trimbox{0cm 0cm 0.3cm 0cm}{
\pgfplotsset{filter discard warning=false}
\pgfplotsset{every axis legend/.append style={
at={(1.55,1)},
anchor=north east}} 

\pgfplotscreateplotcyclelist{whatever}{%
    red,thick,every mark/.append style={fill=blue!80!black},mark=none\\%
    blue,thick,every mark/.append style={fill=red!80!black},mark=none\\%
    green,thick,every mark/.append style={fill=blue!80!black},mark=none\\%
    red,thick,dotted,every mark/.append style={fill=blue!80!black},mark=none\\%
    blue,thick,dotted,every mark/.append style={fill=red!80!black},mark=none\\%
    green,thick,dotted,every mark/.append style={fill=blue!80!black},mark=none\\%
    % gray,thick,dashed,every mark/.append style={fill=red!80!black},mark=none\\%
    % black,thick,every mark/.append style={fill=blue!80!black},mark=none\\%
    % gray,thick,every mark/.append style={fill=red!80!black},mark=none\\%
    % % brown!60!black,every mark/.append style={fill=brown!80!black},mark=triangle*\\%
    % black,mark=star\\%
    % blue,every mark/.append style={fill=blue!80!black},mark=diamond*\\%
    % red,densely dashed,every mark/.append style={solid,fill=red!80!black},mark=*\\%
    % brown!60!black,densely dashed,every mark/.append style={solid,fill=brown!80!black},mark=square*\\%
    % black,densely dashed,every mark/.append style={solid,fill=gray},mark=triangle*\\%
    % blue,densely dashed,mark=star,every mark/.append style=solid\\%
    % red,densely dashed,every mark/.append style={solid,fill=red!80!black},mark=diamond*\\%
    }
    
\begin{tikzpicture}
    \begin{groupplot}[
            group style={
                group size=1 by 1,
                horizontal sep=.05in,
                vertical sep=.05in,
                ylabels at=edge left,
                yticklabels at=edge left,
                xlabels at=edge bottom,
                xticklabels at=edge bottom,
            },
            height=.8in,
            xlabel near ticks,
            ylabel near ticks,
            scale only axis,
            width=0.17*\textwidth,
            xmin=0,
            ymin=0,
            ymax=100,
            %xmax=1.1
        ]

        % \nextgroupplot[
        %     % title=Robustness by demographic group,
        %     ylabel=Certifiable Accuracy,
        %     xmax=0.05,
        %     % xtick={0,0.02,0.04},
        %     % minor xtick={0.01,0.03},
        %     cycle list name=whatever,
        %     xlabel=]
        % \addplot table [x=x,y=2, col sep=comma]{data/cifar10.csv};
        % \addplot table [x=x,y=0, col sep=comma]{data/cifar10.csv};
        % \addplot table [x=x,y=1, col sep=comma]{data/cifar10.csv};
        % \addplot table [x=x,y=2-normal, col sep=comma]{data/cifar10.csv};
        % \addplot table [x=x,y=0-normal, col sep=comma]{data/cifar10.csv};
        % \addplot table [x=x,y=1-normal, col sep=comma]{data/cifar10.csv};
        
        \nextgroupplot[
            xmax=0.02,
            % xtick={0,0.005,0.01,0.015},
            % minor xtick={0.0025,0.0075,0.0125,0.0175},
             cycle list name=whatever,
        ]
        \addplot table [x=x,y=2, col sep=comma]{data/ember.csv};
        \addplot table [x=x,y=0, col sep=comma]{data/ember.csv};
        \addplot table [x=x,y=1, col sep=comma]{data/ember.csv};
        \addplot table [x=x,y=2-normal, col sep=comma]{data/ember.csv};
        \addplot table [x=x,y=0-normal, col sep=comma]{data/ember.csv};
        \addplot table [x=x,y=1-normal, col sep=comma]{data/ember.csv};

\legend{BagFlip, \technique-$200$, \technique-$400$};

%  \node (P1) at (pvgroup c5r1.north east) {};
%     \node (P2) at (pvgroup c5r1.south east) {};
%     \path (P1) -- node[right]{\pgfplotslegendfromname{singlelegend}} (P2);
    
    \end{groupplot}
    % \node at (1.2,-1) {\scriptsize (a) CIFAR10 $\featflipone$};
    \node at (1.6,-1) {\scriptsize (b) EMBER $\featflipone$};
    % \node[draw] at (7.5,1.7) {\scriptsize (c) EMBER $\featflipone$};
    % \node[draw] at (9.8,0.5) {\scriptsize (d) EMBER $\featflipinf$};

\end{tikzpicture}
}
    & \pgfplotsset{filter discard warning=false}
\pgfplotsset{every axis legend/.append style={
at={(1,1)},
anchor=north east}} 

\pgfplotscreateplotcyclelist{whatever}{%
    red,thick,every mark/.append style={fill=blue!80!black},mark=none\\%
    blue,thick,every mark/.append style={fill=red!80!black},mark=none\\%
    green,thick,every mark/.append style={fill=blue!80!black},mark=none\\%
    pink,thick,every mark/.append style={fill=blue!80!black},mark=none\\%
    red,thick,dotted,every mark/.append style={fill=blue!80!black},mark=none\\%
    blue,thick,dotted,every mark/.append style={fill=red!80!black},mark=none\\%
    green,thick,dotted,every mark/.append style={fill=blue!80!black},mark=none\\%
    pink,thick,dotted,every mark/.append style={fill=blue!80!black},mark=none\\%
    % gray,thick,dashed,every mark/.append style={fill=red!80!black},mark=none\\%
    % black,thick,every mark/.append style={fill=blue!80!black},mark=none\\%
    % gray,thick,every mark/.append style={fill=red!80!black},mark=none\\%
    % % brown!60!black,every mark/.append style={fill=brown!80!black},mark=triangle*\\%
    % black,mark=star\\%
    % blue,every mark/.append style={fill=blue!80!black},mark=diamond*\\%
    % red,densely dashed,every mark/.append style={solid,fill=red!80!black},mark=*\\%
    % brown!60!black,densely dashed,every mark/.append style={solid,fill=brown!80!black},mark=square*\\%
    % black,densely dashed,every mark/.append style={solid,fill=gray},mark=triangle*\\%
    % blue,densely dashed,mark=star,every mark/.append style=solid\\%
    % red,densely dashed,every mark/.append style={solid,fill=red!80!black},mark=diamond*\\%
    }
    
\begin{tikzpicture}
    \begin{groupplot}[
            group style={
                group size=1 by 1,
                horizontal sep=.05in,
                vertical sep=.05in,
                ylabels at=edge left,
                yticklabels at=edge left,
                xlabels at=edge bottom,
                xticklabels at=edge bottom,
            },
            height=.8in,
            xlabel near ticks,
            ylabel near ticks,
            scale only axis,
            width=0.35*\textwidth,
            xmin=0,
            ymin=0,
            ymax=100,
            %xmax=1.1
            yticklabels={},
        ]

        \nextgroupplot[
            % title=Robustness by demographic group,
            ylabel=,
            xmax=0.8,
            xtick={0,0.5,1,1.5},
            minor xtick={0.25,0.75,1.25},
            cycle list name=whatever,
            xlabel=Modification Amount $\assumeradii$ (\%)]
        \addplot table [x=x,y=3, col sep=comma]{data/mnist.csv};
        \addplot table [x=x,y=0, col sep=comma]{data/mnist.csv};
        \addplot table [x=x,y=1, col sep=comma]{data/mnist.csv};
        \addplot table [x=x,y=2, col sep=comma]{data/mnist.csv};
        \addplot table [x=x,y=3-normal, col sep=comma]{data/mnist.csv};
        \addplot table [x=x,y=0-normal, col sep=comma]{data/mnist.csv};
        \addplot table [x=x,y=1-normal, col sep=comma]{data/mnist.csv};
        \addplot table [x=x,y=2-normal, col sep=comma]{data/mnist.csv};

\legend{BagFlip, \technique-$600$, \technique-$1200$, \technique-$3000$};

%  \node (P1) at (pvgroup c5r1.north east) {};
%     \node (P2) at (pvgroup c5r1.south east) {};
%     \path (P1) -- node[right]{\pgfplotslegendfromname{singlelegend}} (P2);
    
    \end{groupplot}
    % \node at (2.95,-1) {\scriptsize (a) MNIST $\featflipone$};
    % \node at (4.45,-1) {\scriptsize (b) EMBER $\featflipone$};
    % \node[draw] at (7.5,1.7) {\scriptsize (c) EMBER $\featflipone$};
    % \node[draw] at (9.8,0.5) {\scriptsize (d) EMBER $\featflipinf$};
    \node at (2.9,-1) {\scriptsize (c) MNIST $\featflipone$};

\end{tikzpicture} \\
\end{tabular}

\caption{Comparison to BagFlip on CIFAR10, EMBER and MNIST, showing the normal accuracy (dotted lines) and the certified accuracy (solid lines) at different modification amounts $\assumeradii$. 
% For CIFAR10: $a=50$ and $b=100$. For EMBER: $a=200$ and $b=400$.
}

\label{fig:bd_cifar10_ember}
% \vspace{-2em}
\end{figure*}

% \begin{figure}
% \centering
% \tiny
% \input{figs/mnist_bd.tex}
% \caption{Comparison to BagFlip on MNIST, showing the normal accuracy (dotted lines) and the certified accuracy (solid lines) at different modification amounts $\assumeradii$.}

% \label{fig:bd_mnist}

% \end{figure}

Figure~\ref{fig:bd_cifar10_ember} shows the comparison between \technique and BagFlip on CIFAR10, EMBER, and MNIST. 
\textbf{\technique achieves significantly higher certified accuracy than BagFlip across all modification amounts $\assumeradii$ and normal accuracy similar to  BagFlip for all datasets.}

BagFlip performs poorly on CIFAR10 and EMBER because these two datasets cannot tolerate the high level of noise that the BagFlip algorithm adds to the training data.
\yhmodify{}{A high level of noise is crucial to establish meaningful bounds by BagFlip.}
For example, BagFlip can add 20\% noise to the training data of MNIST, i.e., a feature (pixel) in a training example will be flipped to another value with 20\% probability.
\yhmodify{However, for CIFAR10 and EMBER, this probability has to be decreased to 5\% to maintain normal accuracy.}{
However, for CIFAR10 and EMBER, BagFlip has to reduce this probability to 5\% to maintain normal accuracy, resulting in a low certified accuracy.
}

% Figure~\ref{fig:bd_cifar10_ember}~(c) shows the comparison between \technique and BagFlip on MNIST.
% \textbf{\technique achieves competitive results compared to BagFlip.}
% We find that two approaches have similar normal accuracy.
Figure~\ref{fig:bd_cifar10_ember}~(c) shows the comparison between \technique and BagFlip on MNIST.
\technique-$3000$ achieves higher certified accuracy than BagFlip across all modification amounts $\assumeradii$.
When comparing \technique-$600$ and \technique-$1200$ with BagFlip,
we find that \technique-$600$ and \technique-$1200$ achieve higher certified accuracy than BagFlip when $\assumeradii \in [0,0.23]$ and $\assumeradii \in [0,0.42]$, respectively.

% We argue that the gap of certified accuracy between \technique-$3000$ and BagFlip mainly comes from the different definitions of the perturbation spaces as discussed in Appendix~\ref{sec: appendix_detail_setup}.
% Moreover, the root cause of this difference is owing to the probabilistic nature of BagFlip.

\textbf{\technique is more efficient than BagFlip at computing the certified radius.}
\technique computes the certified radius in constant time via the closed-form solution in the aggregation step.
In our experiment on MNIST, BagFlip requires 8 hours to prepare a lookup table because BagFlip does not have a closed-form solution for computing the certified radius. 
We provide the training time of \technique in Appendix~\ref{sec: appendix_detail_setup}.

\subsection{\technique under Backdoored Data}
\label{sec: rq2}
In Section~\ref{sec: backdoor_badnets}, we assess the efficacy of \technique in image classification using the CIFAR10-02 and MNIST datasets under the BadNets~\cite{badnets}, which introduces patches as triggers to implant backdoors into models.
We compare \technique with Friendly Noise~\cite{friendly_noise}, the state-of-the-art empirical defense against general data poisoning attacks on image classification datasets. Subsequently, we evaluate two certified defenses: DPA, a defense against trigger-less attacks, and PatchGuard++, a defense against evasion patch attacks. Finally, we assess FPA, a certified defense against backdoor attacks.

In Section~\ref{sec: backdoor_xba}, we evaluate five empirical defenses---Isolation Forests~\cite{isolation_forest}, HDBSCAN~\cite{HDBSCAN}, Spectral Signatures~\cite{SpectralSignatures}, MDR~\cite{mdr}, and Friendly Noise---in the context of malware detection using the EMBER dataset under the explanation-guided backdoor attack (XBA)~\cite{backdoor_Malwarepoison}. The first three defenses are proposed as adaptive defenses in XBA, while MDR is currently the state-of-the-art empirical defense against backdoor attacks on the EMBER dataset. Additionally, we assess two certified defenses: DPA, FPA, and CROWN-IBP, a defense against evasion attacks. 
% The evaluation and discussion of FPA are provided in Appendix~\ref{sec: appendix_exp3}.

We do not compare to BagFlip because: (1) BadNets generates large patches beyond BagFlip's defending ability, resulting in a certified accuracy of zero, and (2) BagFlip's certified accuracy on EMBER is poor, as shown in Figure~\ref{fig:bd_cifar10_ember}~(b). 

\begin{tcolorbox}[sharp corners, boxsep=0pt,left=3pt,right=3pt,top=2pt,bottom=2pt]
\paragraph{Summary of the results}
\technique reduces the ASR of the victim model (NoDef) on backdoored test sets from an average of 90.24\% to 0.67\% under BadNets and from 66.37\% to 2.19\% under XBA, while other approaches fail to defend against the backdoor attacks, with the exception of FPA, which performs well under the XBA attack.
% Being the most conservative, \technique has the highest abstention rate.
\end{tcolorbox}
% \vspace{-2em}

% \input{tables/ember_bd_empirical.tex}

% \begin{figure}[t]
%     \input{figs/malware_res_chart.tex}
%     \caption{Results of \technique, DPA, CROWN-IBP (C-IBP), and vanilla model without defense (NoDef) trained on three poisoned EMBER datasets when evaluated on (a) the malware test set with malicious triggers and (b) the (original) malware test set without malicious triggers. We note that NoDef does not have abstention rates because it does not use any defense. }
%     \label{fig: malware_res_chart1}
%     \vspace{-1em}
% \end{figure}

% \begin{figure}[t]
%     \input{figs/malware_res_chart1.tex}
%     \caption{Results of \technique, DPA, C-IBP, and NoDef when evaluated on the (original) malware test set without malicious triggers.}
%     \label{fig: malware_res_chart2}
%     \vspace{-1em}
% \end{figure}
\begin{table*}
    \centering
    \vspace{-1em}
    \caption{Results on poisoned dataset generated by BadNets and evaluated on test sets with triggers and clean test sets.
    We report the standard error of the mean in parentheses.
    We note that NoDef and other empirical methods do not have abstention rates.}
    \vspace{0.5em}
    \resizebox{0.7\textwidth}{!}{
    \setlength{\tabcolsep}{2pt}
    \renewcommand{\arraystretch}{1}
    \small
    \begin{tabular}{lllllll}
    \toprule
        && \multicolumn{3}{c}{Test set with triggers}                 & \multicolumn{2}{c}{Clean test set}
         \\ 
 \cmidrule(lr){3-5} \cmidrule(lr){6-7} 
        &Approaches & ASR ($\downarrow$) & 
        Accuracy ($\uparrow$) &
        Abstention Rate & Accuracy ($\uparrow$) & Abstention Rate
        \\ \midrule
        \multirow{6}{*}{\rotatebox[origin=c]{90}{CIFAR10-02}} & NoDef & 95.54\% (5.32) & 
        ~~4.46\% (5.32) & 
        N/A & 89.74\% (0.90) & N/A\\
        & Friendly Noise & 16.74\% (2.25) & 
        83.26\% (2.25) & 
        N/A & 86.02\% (0.25) & N/A \\ 
        & PatchGuard++ & ~~1.64\% (0.47) & 
        ~~0.30\% (0.37) & 
        98.07\% (0.50) & 77.18\% (2.16) & 18.33\% (2.37) \\ 
        & DPA & ~~6.21\% & 24.60\% & 69.19\% & 70.00\% & 23.45\% \\ 
        & FPA & ~~0.00\% & ~~0.00\% & 100.0\% & ~~0.00\% & 100.0\% \\ 
        & \cellcolor{gray!30}\technique & \cellcolor{gray!30}~~0.87\% & \cellcolor{gray!30}22.86\% & \cellcolor{gray!30}77.27\% & \cellcolor{gray!30}69.60\% & \cellcolor{gray!30}23.95\%  \\ 
        \midrule
        \multirow{6}{*}{\rotatebox[origin=c]{90}{MNIST}} & NoDef & 84.93\% (6.85) & 
        15.07\% (6.85) & 
        N/A & 92.61\% (2.79) & N/A\\
        & Friendly Noise & 11.41\% (5.32) & 
        88.59\% (5.32) & 
        N/A & 85.33\% (4.09) & N/A \\ 
        & PatchGuard++ & ~~0.76\% (0.45) & 
        ~~4.77\% (0.80) & 
        94.47\% (1.07) & 38.71\% (3.77) & 60.84\% (3.45) \\ 
        & DPA & ~~0.46\% & 80.93\% & 18.61\% & 74.92\% & 20.31\% \\ 
        & FPA & ~~0.00\% & ~~0.00\% & 100.0\% & ~~0.00\% & 100.0\% \\ 
        & \cellcolor{gray!30}\technique & \cellcolor{gray!30}~~0.46\% & \cellcolor{gray!30}80.93\% & \cellcolor{gray!30}18.61\% & \cellcolor{gray!30}74.92\% & \cellcolor{gray!30}20.31\%  \\ 
        \bottomrule
    \end{tabular}
    }
    \label{tab:badnet_res}
    \vspace{-1em}
\end{table*}

\subsubsection{Experiments with BadNets}
\label{sec: backdoor_badnets}
\paragraph{Setup}
We inject backdoors into 0.2\% of the CIFAR10-02 and MNIST training sets using BadNets. 
For the CIFAR10-02 dataset, a $8 \times 8$ backdoored patch is added to images labeled with 1, aiming to mislead the victim model into predicting these images as label 0 instead of 1. 
For the MNIST dataset, nine different backdoored patches with sizes $4 \times 4$ are added at various locations to images labeled with 1-9. The objective is to deceive the victim model into predicting the backdoored images as the digit 0.
Note that we do not change labels of the poisoned training data.

Our hyperparameter tuning strategy of Friendly Noise follows the original paper. 
The number of partitions in FPA is set to 256 and 56 for CIFAR10-02 and MNIST datasets, respectively.
For PatchGuard++ and \technique, the hyperparameter $\tau$, which controls the trade-off between ASR and abstention rate, is set to maximize the difference between accuracy and ASR (refer to Appendix~\ref{sec: appendix_rq2_detail}). 
\yhmodify{The approaches are executed five times, and the average results are reported, except for ensemble-based methods \technique, DPA, and FPA, which inherently exhibit low variance.}{
We run ensemble-based methods \technique, DPA, and FPA only once because they inherently exhibit low variance. For other approaches, we run them five times and report the average results.}
% \loris{so you run once? or report something else?}

% Table~\ref{tab:badnet_res} reports the results of all approaches, including the victim classifier with no defense mechanism (NoDef), on the test sets with triggers and clean test sets.

\paragraph{Comparison to Friendly Noise} Friendly Noise, the state-of-the-art empirical defense, fails to defend against the BadNets attack, showing high ASRs of 16.74\% and 11.41\% on CIFAR10-02 and MNIST datasets, respectively. In contrast, \technique achieves much lower ASRs of 0.87\% and 0.46\%.
Compared to \technique, Friendly Noise achieves higher accuracies on both test sets as \technique is a certified defense and has high abstention rates.

\paragraph{Comparison to DPA and PatchGuard++}
DPA fails to defend against the BadNets attack on the CIFAR10-02 dataset with an ASR of 6.21\%.
The result of DPA is the same as \technique in the MNIST dataset because DPA is equivalent to \technique when its hyperparameter $\tau$ is set to $1$.
PatchGuard++ shows high ASRs across all values of $\tau \in [0, 1]$, with optimal performance achieved when setting $\tau$ to 0, leading to high abstention rates.

\paragraph{Comparison to FPA} FPA is unsuitable for defending against the BadNets attack due to the large patch sizes, limiting each classifier in FPA to consider only 4 and 14 pixels for each input in CIFAR10-02 and MNIST datasets.
This limited visibility of pixels results in each classifier in FPA merely guessing predictions.
Additionally, FPA cannot defend against dynamic backdoor attacks that can place patches at different locations. 

\begin{table*}
    \centering
    \vspace{-1em}
    \caption{Results on poisoned dataset $\poisoned{D}_3$ when evaluated on the malware test set with triggers and the clean test set.}
    \vspace{0.5em}
    \resizebox{0.7\textwidth}{!}{
    \setlength{\tabcolsep}{2pt}
    \renewcommand{\arraystretch}{1}
    \small
    \begin{tabular}{llllll}
    \toprule
        & \multicolumn{3}{c}{Malware test set with triggers}                 & \multicolumn{2}{c}{Clean test set}
        \\ 
 \cmidrule(lr){2-4} \cmidrule(lr){5-6} 
        Approaches & ASR ($\downarrow$) & Accuracy ($\uparrow$) & Abstention Rate & Accuracy  ($\uparrow$) & Abstention Rate 
        \\ \midrule
        NoDef & 67.16\% (11.21) & 32.84\% (11.21) & N/A  & 98.37\% (0.28) & N/A\\ 
        Isolation Forest & 28.74\% (~~8.71) & 71.26\% (~~8.71) & N/A  & 94.16\% (0.24) & N/A  \\ 
        HDBSCAN & 63.47\% (11.85) & 36.53\% (11.85) & N/A  & 98.11\% (0.37) & N/A  \\ 
        Spectral Signature & 67.07\% (16.63) & 32.93\% (16.63) & N/A  & 98.11\% (0.40) & N/A  \\ 
        MDR & 63.56\% (10.98) & 36.44\% (10.98) & N/A & 98.27\% (0.21) & N/A  \\ 
        Friendly Noise & 58.76\% (~~1.57) & 41.24\% (~~1.57) & N/A & 95.09\% (0.58) & N/A \\ 
        \rowcolor{gray!30}\technique-Empirical & 25.47\% & 74.53\% & N/A & 89.15\% & N/A \\ 
        CROWN-IBP & ~~6.64\% (~~2.25) & 28.87\% (~~2.23) & 64.49\% (2.29) & 62.04\% (0.85) & 32.27\% (1.26)  \\ 
        DPA & 33.91\% & 41.89\% & 24.20\% & 79.06\% & 5.20\% \\ 
        FPA & ~~0.72\% & 34.28\% & 65.00\% & 76.41\% & 23.38\% \\ 
        \rowcolor{gray!30}\technique & ~~2.19\% & 29.46\% & 68.35\% & 42.44\% & 56.42\% \\ 
        \bottomrule
    \end{tabular}
    }
    \label{tab:xba_res}
    \vspace{-1em}
\end{table*}

\subsubsection{Experiments with XBA}
\label{sec: backdoor_xba}
\paragraph{Setup}
We use XBA to backdoor 0.1\% of the training set and add triggers into the malware in the test set.
We aim to fool the victim model to predict the malware with malicious triggers as non-malware.  
We generate a poisoned dataset $\poisoned{D}_3$ and its corresponding test set with triggers by perturbation $\featflipthree$, which allows the attacker to modify up to three features in an example. 
As shown in Table~\ref{tab:xba_res}, XBA achieves a 67.16\% ASR with only three features modified.
% \loris{is 3 good?}

% We report the results of all approaches and the victim classifier with no defense (NoDef) mechanism in place on the malware test sets with triggers.
% The results on test sets without triggers, i.e., the original malware test set and the original non-malware test sets can be found in Appendix~\ref{sec: appendix_rq2_detail}.
% ASR on malware is a much more critical metric than the ASR on non-malware, because the former shows how many pieces of malware can bypass the classifier.

Appendix~\ref{sec: appendix_rq2_detail} presents the hyperparameter settings of the empirical approaches.
For \technique and DPA, we set the number of partitions $n$ to 3000 and present their results for the modification amount $\assumeradii = 0.1\%$. 
As CROWN-IBP does not consider $\assumeradii$, we show its results against the perturbation $\featflipthree$ regardless of $\assumeradii$. 

% Table~\ref{tab:xba_res} shows the ASR, accuracy, and abstention rate of all approaches on the malware test set with triggers and the accuracy and abstention rate on the clean test set.

\paragraph{Comparison to Empirical Defenses}
\technique can defend against the backdoor attack on the EMBER dataset, but Isolation Forest, HDBSCAN, Spectral Signature, and MDR fail to defend against the attack.
Table~\ref{tab:xba_res} shows that \technique successfully reduces the ASR of the victim model from 67.16\% to 2.19\% on the malware test set with triggers.
\technique has the lowest ASR compared to other empirical defenses, in which the best empirical defense, Isolation Forest, has an ASR of 28.74\%. 
We found that these empirical approaches struggled to filter out poisoning examples in the training set due to the small amount of poisoning---only 0.1\%. We report their precision and recall of filtering in Appendix~\ref{sec: res_filtering}.

\technique has a high abstention rate because \technique is a certified defense against backdoor attacks.
However, even if we make \technique an empirical defense by producing an output even when certification fails, we find \technique-Empirical still achieves an ASR of 25.47\% and an accuracy of 74.53\%, which still outperforms the best empirical approach Isolation Forest, with an ASR 28.74\%.

\paragraph{Comparison to DPA and CROWN-IBP} 
The ASR of DPA and CROWN-IBP on the malware test set with triggers are $33.91\%$ and $6.64\%$ meaning that many malware with triggers can bypass their defenses.

\paragraph{Comparison to FPA}
FPA defends against backdoor attacks by adopting a feature partitioning-approach, making it effective against attacks that introduce a fixed and small trigger across all examples, especially on tabular datasets where useful information is preserved after partitioning. 
As a result, FPA achieves a lower ASR of 0.72\% than \technique on the EMBER dataset under the XBA attack.

However, FPA has limitations when defending against attacks with medium-size dynamic triggers, such as the BadNets attack employed in Section~\ref{sec: backdoor_badnets}, or on image datasets where features are significantly disrupted after partitioning. 
Additionally, FPA cannot defend against attacks that involve modifications to training labels or the insertion/removal of training examples.
In contrast, \technique does not have these limitations because its perturbation space captures all such data poisoning attacks. 
Consequently, \technique successfully defends against both BadNets and XBA attacks. 
% Further details and comparisons between FPA and \technique are provided in Appendix~\ref{sec: appendix_exp3}, where we extend XBA to create another poisoned dataset by introducing different triggers on training and test examples.

In conclusion, while FPA demonstrates effectiveness on fixed and small triggers, particularly on tabular datasets like EMBER under the XBA attack, it faces challenges against dynamic triggers and disruptions in image datasets. 
On the other hand, \technique is more versatile against a broader range of data poisoning attacks. 
\section{Conclusion, Limitations, and Future Work}
\label{sec: conclusion}
We presented \technique, a deterministic certified approach  to effectively and efficiently defend against backdoor attacks. 
We foresee many future improvements to \technique. 
First, \technique generates small certified radii for large datasets such as CIFAR10 and EMBER. 
Thus, \technique prevents attackers from using a small amount of poison to make their attacks more difficult to detect. 
Furthermore, as shown in our experiments and \citet{pickyourpoison}, empirical approaches alone cannot defend against backdoor attacks that use small fragments of poisoned examples. 
We argue that \technique can complement empirical defenses in the real world.
Second, \technique equipped with CROWN-IBP currently only works with simple neural networks and cannot be extended to large datasets like ImageNet. 
However, when equipped with PatchGuard++, \technique can work with more complex model structures like BagNet, whose size is similar to a ResNet-50 model.
Furthermore, \technique with CROWN-IBP can be extended to more complex models and datasets in the future, as witnessed by the growth of robust training~\cite{smallbox} and evasion certification techniques being extended to these models and datasets. Nevertheless, directly applying these techniques to \technique is currently computationally infeasible, as it takes 42 minutes for \technique to use $\alpha\beta$-CROWN~\cite{betacrown} to certify one input in TinyImageNet for one hundred ResNet models. 
Sharing the intermediary certification results among different models~\cite{shareproof, shared_sat, shared_split} can significantly improve the efficiency of \technique, and we leave this as future work.
Third, we adopt the idea of deep partition aggregation (DPA) to design the partition and aggregation steps in \technique. We can improve these steps by using finite aggregation~\cite{dpa_improve} and run-off election~\cite{runoff}, which extends DPA and gives higher certified accuracy.

\newpage
\section{Impact Statements}
This paper introduces \technique, a novel certified defense against backdoor attacks in neural networks. 

\paragraph{Positive Impacts}
\technique positively contributes to the security and integrity of machine learning by providing a certified defense against backdoor attacks, particularly in critical applications such as malware detection.
The partitioning and ensemble strategy employed by \technique has the potential to address the fairness concern that training data miss the data points from under-represented groups. 
As \technique can define the perturbation space concerning training examples insertion and deletion, we can ensure that the model prediction will not be affected by the historical biases present in the training data or let the model abstain from making biased predictions.

We outline the potential negative societal impacts and mitigation strategies as follows.
\paragraph{Advanced Availability Attacks} While \technique offers a certified defense, ensuring that adversaries cannot manipulate the model's prediction to a specific target class, it is important to consider that adversaries may develop more complex availability attacks. These attacks could potentially cause \technique to abstain from making predictions, undermining the system's usability.
We suggest incorporating fallback mechanisms or human-in-the-loop systems to handle cases where the model abstains, ensuring that critical decisions are made promptly.
    
\paragraph{Resource Accessibility}  Moreover, the increased complexity and computational resources required for certified defenses may limit the accessibility of secure machine learning models to organizations with fewer resources.
    To address this, we advocate for developing more efficient algorithms and providing cloud-based services that offer certified defenses at a lower cost.

In summary, \technique represents a significant advancement in machine learning security. However, it is crucial to remain cognizant of the evolving landscape of adversarial threats and the need for equitable access to state-of-the-art defense technologies.

\bibliography{main}

\begin{thebibliography}{56}
\providecommand{\natexlab}[1]{#1}
\providecommand{\url}[1]{\texttt{#1}}
\expandafter\ifx\csname urlstyle\endcsname\relax
  \providecommand{\doi}[1]{doi: #1}\else
  \providecommand{\doi}{doi: \begingroup \urlstyle{rm}\Url}\fi

\bibitem[Anderson and Roth(2018)]{ember}
Hyrum~S. Anderson and Phil Roth.
\newblock {EMBER:} an open dataset for training static {PE} malware machine
  learning models.
\newblock \emph{CoRR}, abs/1804.04637, 2018.
\newblock URL \url{http://arxiv.org/abs/1804.04637}.

\bibitem[Brendel and Bethge(2019)]{bagnet}
Wieland Brendel and Matthias Bethge.
\newblock Approximating cnns with bag-of-local-features models works
  surprisingly well on imagenet.
\newblock In \emph{7th International Conference on Learning Representations,
  {ICLR} 2019, New Orleans, LA, USA, May 6-9, 2019}. OpenReview.net, 2019.
\newblock URL \url{https://openreview.net/forum?id=SkfMWhAqYQ}.

\bibitem[Brown et~al.(2017)Brown, Man{\'{e}}, Roy, Abadi, and Gilmer]{advpatch}
Tom~B. Brown, Dandelion Man{\'{e}}, Aurko Roy, Mart{\'{\i}}n Abadi, and Justin
  Gilmer.
\newblock Adversarial patch.
\newblock \emph{CoRR}, abs/1712.09665, 2017.
\newblock URL \url{http://arxiv.org/abs/1712.09665}.

\bibitem[Chen et~al.(2020)Chen, Li, Wu, Sheng, and Li]{framework_def}
Ruoxin Chen, Jie Li, Chentao Wu, Bin Sheng, and Ping Li.
\newblock A framework of randomized selection based certified defenses against
  data poisoning attacks.
\newblock \emph{CoRR}, abs/2009.08739, 2020.
\newblock URL \url{https://arxiv.org/abs/2009.08739}.

\bibitem[Chen et~al.(2022)Chen, Li, Li, Yan, and Wu]{collective}
Ruoxin Chen, Zenan Li, Jie Li, Junchi Yan, and Chentao Wu.
\newblock On collective robustness of bagging against data poisoning.
\newblock In Kamalika Chaudhuri, Stefanie Jegelka, Le~Song, Csaba
  Szepesv{\'{a}}ri, Gang Niu, and Sivan Sabato, editors, \emph{International
  Conference on Machine Learning, {ICML} 2022, 17-23 July 2022, Baltimore,
  Maryland, {USA}}, volume 162 of \emph{Proceedings of Machine Learning
  Research}, pages 3299--3319. {PMLR}, 2022.
\newblock URL \url{https://proceedings.mlr.press/v162/chen22k.html}.

\bibitem[Cohen et~al.(2019)Cohen, Rosenfeld, and Kolter]{randomizesmoothing}
Jeremy~M. Cohen, Elan Rosenfeld, and J.~Zico Kolter.
\newblock Certified adversarial robustness via randomized smoothing.
\newblock In Kamalika Chaudhuri and Ruslan Salakhutdinov, editors,
  \emph{Proceedings of the 36th International Conference on Machine Learning,
  {ICML} 2019, 9-15 June 2019, Long Beach, California, {USA}}, volume~97 of
  \emph{Proceedings of Machine Learning Research}, pages 1310--1320. {PMLR},
  2019.
\newblock URL \url{http://proceedings.mlr.press/v97/cohen19c.html}.

\bibitem[Drews et~al.(2020)Drews, Albarghouthi, and D'Antoni]{decisiontree}
Samuel Drews, Aws Albarghouthi, and Loris D'Antoni.
\newblock Proving data-poisoning robustness in decision trees.
\newblock In Alastair~F. Donaldson and Emina Torlak, editors, \emph{Proceedings
  of the 41st {ACM} {SIGPLAN} International Conference on Programming Language
  Design and Implementation, {PLDI} 2020, London, UK, June 15-20, 2020}, pages
  1083--1097. {ACM}, 2020.
\newblock \doi{10.1145/3385412.3385975}.
\newblock URL \url{https://doi.org/10.1145/3385412.3385975}.

\bibitem[Dvijotham et~al.(2020)Dvijotham, Hayes, Balle, Kolter, Qin,
  Gy{\"{o}}rgy, Xiao, Gowal, and Kohli]{divergences}
Krishnamurthy~(Dj) Dvijotham, Jamie Hayes, Borja Balle, J.~Zico Kolter, Chongli
  Qin, Andr{\'{a}}s Gy{\"{o}}rgy, Kai Xiao, Sven Gowal, and Pushmeet Kohli.
\newblock A framework for robustness certification of smoothed classifiers
  using f-divergences.
\newblock In \emph{8th International Conference on Learning Representations,
  {ICLR} 2020, Addis Ababa, Ethiopia, April 26-30, 2020}. OpenReview.net, 2020.
\newblock URL \url{https://openreview.net/forum?id=SJlKrkSFPH}.

\bibitem[Fischer et~al.(2022)Fischer, Sprecher, Dimitrov, Singh, and
  Vechev]{shareproof}
Marc Fischer, Christian Sprecher, Dimitar~I. Dimitrov, Gagandeep Singh, and
  Martin~T. Vechev.
\newblock Shared certificates for neural network verification.
\newblock In Sharon Shoham and Yakir Vizel, editors, \emph{Computer Aided
  Verification - 34th International Conference, {CAV} 2022, Haifa, Israel,
  August 7-10, 2022, Proceedings, Part {I}}, volume 13371 of \emph{Lecture
  Notes in Computer Science}, pages 127--148. Springer, 2022.
\newblock \doi{10.1007/978-3-031-13185-1\_7}.
\newblock URL \url{https://doi.org/10.1007/978-3-031-13185-1\_7}.

\bibitem[Geiping et~al.(2021)Geiping, Fowl, Somepalli, Goldblum, Moeller, and
  Goldstein]{heu_advtrain}
Jonas Geiping, Liam Fowl, Gowthami Somepalli, Micah Goldblum, Michael Moeller,
  and Tom Goldstein.
\newblock What doesn't kill you makes you robust(er): Adversarial training
  against poisons and backdoors.
\newblock \emph{CoRR}, abs/2102.13624, 2021.
\newblock URL \url{https://arxiv.org/abs/2102.13624}.

\bibitem[Gu et~al.(2017)Gu, Dolan{-}Gavitt, and Garg]{badnets}
Tianyu Gu, Brendan Dolan{-}Gavitt, and Siddharth Garg.
\newblock Badnets: Identifying vulnerabilities in the machine learning model
  supply chain.
\newblock \emph{CoRR}, abs/1708.06733, 2017.
\newblock URL \url{http://arxiv.org/abs/1708.06733}.

\bibitem[Hammoudeh and Lowd(2023)]{fpa}
Zayd Hammoudeh and Daniel Lowd.
\newblock Feature partition aggregation: {A} fast certified defense against a
  union of sparse adversarial attacks.
\newblock \emph{CoRR}, abs/2302.11628, 2023.
\newblock \doi{10.48550/arXiv.2302.11628}.
\newblock URL \url{https://doi.org/10.48550/arXiv.2302.11628}.

\bibitem[Jia et~al.(2020)Jia, Cao, and Gong]{certkrNN}
Jinyuan Jia, Xiaoyu Cao, and Neil~Zhenqiang Gong.
\newblock Certified robustness of nearest neighbors against data poisoning
  attacks.
\newblock \emph{CoRR}, abs/2012.03765, 2020.
\newblock URL \url{https://arxiv.org/abs/2012.03765}.

\bibitem[Jia et~al.(2021)Jia, Cao, and Gong]{bagging}
Jinyuan Jia, Xiaoyu Cao, and Neil~Zhenqiang Gong.
\newblock Intrinsic certified robustness of bagging against data poisoning
  attacks.
\newblock In \emph{Thirty-Fifth {AAAI} Conference on Artificial Intelligence,
  {AAAI} 2021, Thirty-Third Conference on Innovative Applications of Artificial
  Intelligence, {IAAI} 2021, The Eleventh Symposium on Educational Advances in
  Artificial Intelligence, {EAAI} 2021, Virtual Event, February 2-9, 2021},
  pages 7961--7969. {AAAI} Press, 2021.
\newblock URL \url{https://ojs.aaai.org/index.php/AAAI/article/view/16971}.

\bibitem[Katz et~al.(2019)Katz, Huang, Ibeling, Julian, Lazarus, Lim, Shah,
  Thakoor, Wu, Zeljic, Dill, Kochenderfer, and Barrett]{Marabou}
Guy Katz, Derek~A. Huang, Duligur Ibeling, Kyle Julian, Christopher Lazarus,
  Rachel Lim, Parth Shah, Shantanu Thakoor, Haoze Wu, Aleksandar Zeljic,
  David~L. Dill, Mykel~J. Kochenderfer, and Clark~W. Barrett.
\newblock The marabou framework for verification and analysis of deep neural
  networks.
\newblock In Isil Dillig and Serdar Tasiran, editors, \emph{Computer Aided
  Verification - 31st International Conference, {CAV} 2019, New York City, NY,
  USA, July 15-18, 2019, Proceedings, Part {I}}, volume 11561 of \emph{Lecture
  Notes in Computer Science}, pages 443--452. Springer, 2019.
\newblock \doi{10.1007/978-3-030-25540-4\_26}.
\newblock URL \url{https://doi.org/10.1007/978-3-030-25540-4\_26}.

\bibitem[Koh et~al.(2022)Koh, Steinhardt, and Liang]{break_sanitization}
Pang~Wei Koh, Jacob Steinhardt, and Percy Liang.
\newblock Stronger data poisoning attacks break data sanitization defenses.
\newblock \emph{Mach. Learn.}, 111\penalty0 (1):\penalty0 1--47, 2022.
\newblock \doi{10.1007/s10994-021-06119-y}.
\newblock URL \url{https://doi.org/10.1007/s10994-021-06119-y}.

\bibitem[Lee et~al.(2019)Lee, Yuan, Chang, and Jaakkola]{tightflipbound}
Guang{-}He Lee, Yang Yuan, Shiyu Chang, and Tommi~S. Jaakkola.
\newblock Tight certificates of adversarial robustness for randomly smoothed
  classifiers.
\newblock In Hanna~M. Wallach, Hugo Larochelle, Alina Beygelzimer, Florence
  d'Alch{\'{e}}{-}Buc, Emily~B. Fox, and Roman Garnett, editors, \emph{Advances
  in Neural Information Processing Systems 32: Annual Conference on Neural
  Information Processing Systems 2019, NeurIPS 2019, December 8-14, 2019,
  Vancouver, BC, Canada}, pages 4911--4922, 2019.
\newblock URL
  \url{https://proceedings.neurips.cc/paper/2019/hash/fa2e8c4385712f9a1d24c363a2cbe5b8-Abstract.html}.

\bibitem[Levine and Feizi(2021)]{partitioning}
Alexander Levine and Soheil Feizi.
\newblock Deep partition aggregation: Provable defenses against general
  poisoning attacks.
\newblock In \emph{9th International Conference on Learning Representations,
  {ICLR} 2021, Virtual Event, Austria, May 3-7, 2021}. OpenReview.net, 2021.
\newblock URL \url{https://openreview.net/forum?id=YUGG2tFuPM}.

\bibitem[Liu et~al.(2008)Liu, Ting, and Zhou]{isolation_forest}
Fei~Tony Liu, Kai~Ming Ting, and Zhi{-}Hua Zhou.
\newblock Isolation forest.
\newblock In \emph{Proceedings of the 8th {IEEE} International Conference on
  Data Mining {(ICDM} 2008), December 15-19, 2008, Pisa, Italy}, pages
  413--422. {IEEE} Computer Society, 2008.
\newblock \doi{10.1109/ICDM.2008.17}.
\newblock URL \url{https://doi.org/10.1109/ICDM.2008.17}.

\bibitem[Liu et~al.(2018)Liu, Dolan{-}Gavitt, and Garg]{heu_finetune}
Kang Liu, Brendan Dolan{-}Gavitt, and Siddharth Garg.
\newblock Fine-pruning: Defending against backdooring attacks on deep neural
  networks.
\newblock In Michael Bailey, Thorsten Holz, Manolis Stamatogiannakis, and
  Sotiris Ioannidis, editors, \emph{Research in Attacks, Intrusions, and
  Defenses - 21st International Symposium, {RAID} 2018, Heraklion, Crete,
  Greece, September 10-12, 2018, Proceedings}, volume 11050 of \emph{Lecture
  Notes in Computer Science}, pages 273--294. Springer, 2018.
\newblock \doi{10.1007/978-3-030-00470-5\_13}.
\newblock URL \url{https://doi.org/10.1007/978-3-030-00470-5\_13}.

\bibitem[Liu et~al.(2022)Liu, Yang, and Mirzasoleiman]{friendly_noise}
Tian~Yu Liu, Yu~Yang, and Baharan Mirzasoleiman.
\newblock Friendly noise against adversarial noise: {A} powerful defense
  against data poisoning attack.
\newblock In \emph{NeurIPS}, 2022.
\newblock URL
  \url{http://papers.nips.cc/paper\_files/paper/2022/hash/4e81308aa2eb8e2e4eccf122d4827af7-Abstract-Conference.html}.

\bibitem[Lukas and Kerschbaum(2023)]{pickyourpoison}
Nils Lukas and Florian Kerschbaum.
\newblock Pick your poison: Undetectability versus robustness in data poisoning
  attacks against deep image classification.
\newblock \emph{CoRR}, abs/2305.09671, 2023.
\newblock \doi{10.48550/ARXIV.2305.09671}.
\newblock URL \url{https://doi.org/10.48550/arXiv.2305.09671}.

\bibitem[Ma et~al.(2019)Ma, Zhu, and Hsu]{differential}
Yuzhe Ma, Xiaojin Zhu, and Justin Hsu.
\newblock Data poisoning against differentially-private learners: Attacks and
  defenses.
\newblock In Sarit Kraus, editor, \emph{Proceedings of the Twenty-Eighth
  International Joint Conference on Artificial Intelligence, {IJCAI} 2019,
  Macao, China, August 10-16, 2019}, pages 4732--4738. ijcai.org, 2019.
\newblock \doi{10.24963/ijcai.2019/657}.
\newblock URL \url{https://doi.org/10.24963/ijcai.2019/657}.

\bibitem[Madry et~al.(2018)Madry, Makelov, Schmidt, Tsipras, and Vladu]{pgd}
Aleksander Madry, Aleksandar Makelov, Ludwig Schmidt, Dimitris Tsipras, and
  Adrian Vladu.
\newblock Towards deep learning models resistant to adversarial attacks.
\newblock In \emph{6th International Conference on Learning Representations,
  {ICLR} 2018, Vancouver, BC, Canada, April 30 - May 3, 2018, Conference Track
  Proceedings}. OpenReview.net, 2018.
\newblock URL \url{https://openreview.net/forum?id=rJzIBfZAb}.

\bibitem[Meyer et~al.(2021)Meyer, Albarghouthi, and
  D'Antoni]{prog_decisiontree}
Anna~P. Meyer, Aws Albarghouthi, and Loris D'Antoni.
\newblock Certifying robustness to programmable data bias in decision trees.
\newblock In Marc'Aurelio Ranzato, Alina Beygelzimer, Yann~N. Dauphin, Percy
  Liang, and Jennifer~Wortman Vaughan, editors, \emph{Advances in Neural
  Information Processing Systems 34: Annual Conference on Neural Information
  Processing Systems 2021, NeurIPS 2021, December 6-14, 2021, virtual}, pages
  26276--26288, 2021.
\newblock URL
  \url{https://proceedings.neurips.cc/paper/2021/hash/dcf531edc9b229acfe0f4b87e1e278dd-Abstract.html}.

\bibitem[M{\"{u}}ller et~al.(2022)M{\"{u}}ller, Eckert, Fischer, and
  Vechev]{smallbox}
Mark~Niklas M{\"{u}}ller, Franziska Eckert, Marc Fischer, and Martin~T. Vechev.
\newblock Certified training: Small boxes are all you need.
\newblock \emph{CoRR}, abs/2210.04871, 2022.
\newblock \doi{10.48550/arXiv.2210.04871}.
\newblock URL \url{https://doi.org/10.48550/arXiv.2210.04871}.

\bibitem[Murtagh and Contreras(2012)]{HDBSCAN}
Fionn Murtagh and Pedro Contreras.
\newblock Algorithms for hierarchical clustering: an overview.
\newblock \emph{WIREs Data Mining Knowl. Discov.}, 2\penalty0 (1):\penalty0
  86--97, 2012.
\newblock \doi{10.1002/widm.53}.
\newblock URL \url{https://doi.org/10.1002/widm.53}.

\bibitem[Nguyen et~al.(2022)Nguyen, Lai, Phan, and Thai]{xrand}
Truc D.~T. Nguyen, Phung Lai, NhatHai Phan, and My~T. Thai.
\newblock Xrand: Differentially private defense against explanation-guided
  attacks.
\newblock \emph{CoRR}, abs/2212.04454, 2022.
\newblock \doi{10.48550/arXiv.2212.04454}.
\newblock URL \url{https://doi.org/10.48550/arXiv.2212.04454}.

\bibitem[Qi et~al.(2021)Qi, Yao, Xu, Liu, and Sun]{backdoor_NLP}
Fanchao Qi, Yuan Yao, Sophia Xu, Zhiyuan Liu, and Maosong Sun.
\newblock Turn the combination lock: Learnable textual backdoor attacks via
  word substitution.
\newblock In Chengqing Zong, Fei Xia, Wenjie Li, and Roberto Navigli, editors,
  \emph{Proceedings of the 59th Annual Meeting of the Association for
  Computational Linguistics and the 11th International Joint Conference on
  Natural Language Processing, {ACL/IJCNLP} 2021, (Volume 1: Long Papers),
  Virtual Event, August 1-6, 2021}, pages 4873--4883. Association for
  Computational Linguistics, 2021.
\newblock \doi{10.18653/v1/2021.acl-long.377}.
\newblock URL \url{https://doi.org/10.18653/v1/2021.acl-long.377}.

\bibitem[Rezaei et~al.(2023)Rezaei, Banihashem, Chegini, and Feizi]{runoff}
Keivan Rezaei, Kiarash Banihashem, Atoosa~Malemir Chegini, and Soheil Feizi.
\newblock Run-off election: Improved provable defense against data poisoning
  attacks.
\newblock In Andreas Krause, Emma Brunskill, Kyunghyun Cho, Barbara Engelhardt,
  Sivan Sabato, and Jonathan Scarlett, editors, \emph{International Conference
  on Machine Learning, {ICML} 2023, 23-29 July 2023, Honolulu, Hawaii, {USA}},
  volume 202 of \emph{Proceedings of Machine Learning Research}, pages
  29030--29050. {PMLR}, 2023.
\newblock URL \url{https://proceedings.mlr.press/v202/rezaei23a.html}.

\bibitem[Rosenfeld et~al.(2020)Rosenfeld, Winston, Ravikumar, and
  Kolter]{labelflip}
Elan Rosenfeld, Ezra Winston, Pradeep Ravikumar, and J.~Zico Kolter.
\newblock Certified robustness to label-flipping attacks via randomized
  smoothing.
\newblock In \emph{Proceedings of the 37th International Conference on Machine
  Learning, {ICML} 2020, 13-18 July 2020, Virtual Event}, volume 119 of
  \emph{Proceedings of Machine Learning Research}, pages 8230--8241. {PMLR},
  2020.
\newblock URL \url{http://proceedings.mlr.press/v119/rosenfeld20b.html}.

\bibitem[Saha et~al.(2020)Saha, Subramanya, and Pirsiavash]{backdoor_hidden}
Aniruddha Saha, Akshayvarun Subramanya, and Hamed Pirsiavash.
\newblock Hidden trigger backdoor attacks.
\newblock In \emph{The Thirty-Fourth {AAAI} Conference on Artificial
  Intelligence, {AAAI} 2020, The Thirty-Second Innovative Applications of
  Artificial Intelligence Conference, {IAAI} 2020, The Tenth {AAAI} Symposium
  on Educational Advances in Artificial Intelligence, {EAAI} 2020, New York,
  NY, USA, February 7-12, 2020}, pages 11957--11965. {AAAI} Press, 2020.
\newblock URL \url{https://ojs.aaai.org/index.php/AAAI/article/view/6871}.

\bibitem[Salem et~al.(2022)Salem, Wen, Backes, Ma, and Zhang]{dynamic_trigger}
Ahmed Salem, Rui Wen, Michael Backes, Shiqing Ma, and Yang Zhang.
\newblock Dynamic backdoor attacks against machine learning models.
\newblock In \emph{7th {IEEE} European Symposium on Security and Privacy,
  EuroS{\&}P 2022, Genoa, Italy, June 6-10, 2022}, pages 703--718. {IEEE},
  2022.
\newblock \doi{10.1109/EuroSP53844.2022.00049}.
\newblock URL \url{https://doi.org/10.1109/EuroSP53844.2022.00049}.

\bibitem[Severi et~al.(2021)Severi, Meyer, Coull, and
  Oprea]{backdoor_Malwarepoison}
Giorgio Severi, Jim Meyer, Scott~E. Coull, and Alina Oprea.
\newblock Explanation-guided backdoor poisoning attacks against malware
  classifiers.
\newblock In Michael Bailey and Rachel Greenstadt, editors, \emph{30th {USENIX}
  Security Symposium, {USENIX} Security 2021, August 11-13, 2021}, pages
  1487--1504. {USENIX} Association, 2021.
\newblock URL
  \url{https://www.usenix.org/conference/usenixsecurity21/presentation/severi}.

\bibitem[Shafahi et~al.(2018)Shafahi, Huang, Najibi, Suciu, Studer, Dumitras,
  and Goldstein]{triggerless_featurecollision}
Ali Shafahi, W.~Ronny Huang, Mahyar Najibi, Octavian Suciu, Christoph Studer,
  Tudor Dumitras, and Tom Goldstein.
\newblock Poison frogs! targeted clean-label poisoning attacks on neural
  networks.
\newblock In Samy Bengio, Hanna~M. Wallach, Hugo Larochelle, Kristen Grauman,
  Nicol{\`{o}} Cesa{-}Bianchi, and Roman Garnett, editors, \emph{Advances in
  Neural Information Processing Systems 31: Annual Conference on Neural
  Information Processing Systems 2018, NeurIPS 2018, December 3-8, 2018,
  Montr{\'{e}}al, Canada}, pages 6106--6116, 2018.
\newblock URL
  \url{https://proceedings.neurips.cc/paper/2018/hash/22722a343513ed45f14905eb07621686-Abstract.html}.

\bibitem[Singh et~al.(2019)Singh, Gehr, P{\"{u}}schel, and Vechev]{deeppoly}
Gagandeep Singh, Timon Gehr, Markus P{\"{u}}schel, and Martin~T. Vechev.
\newblock An abstract domain for certifying neural networks.
\newblock \emph{Proc. {ACM} Program. Lang.}, 3\penalty0 ({POPL}):\penalty0
  41:1--41:30, 2019.
\newblock \doi{10.1145/3290354}.
\newblock URL \url{https://doi.org/10.1145/3290354}.

\bibitem[Tran et~al.(2018)Tran, Li, and Madry]{SpectralSignatures}
Brandon Tran, Jerry Li, and Aleksander Madry.
\newblock Spectral signatures in backdoor attacks.
\newblock In Samy Bengio, Hanna~M. Wallach, Hugo Larochelle, Kristen Grauman,
  Nicol{\`{o}} Cesa{-}Bianchi, and Roman Garnett, editors, \emph{Advances in
  Neural Information Processing Systems 31: Annual Conference on Neural
  Information Processing Systems 2018, NeurIPS 2018, December 3-8, 2018,
  Montr{\'{e}}al, Canada}, pages 8011--8021, 2018.
\newblock URL
  \url{https://proceedings.neurips.cc/paper/2018/hash/280cf18baf4311c92aa5a042336587d3-Abstract.html}.

\bibitem[Turner et~al.(2019)Turner, Tsipras, and
  Madry]{backdoor_labelconsistent}
Alexander Turner, Dimitris Tsipras, and Aleksander Madry.
\newblock Label-consistent backdoor attacks.
\newblock \emph{CoRR}, abs/1912.02771, 2019.
\newblock URL \url{http://arxiv.org/abs/1912.02771}.

\bibitem[Ugare et~al.(2023)Ugare, Banerjee, Misailovic, and
  Singh]{shared_split}
Shubham Ugare, Debangshu Banerjee, Sasa Misailovic, and Gagandeep Singh.
\newblock Incremental verification of neural networks.
\newblock \emph{CoRR}, abs/2304.01874, 2023.
\newblock \doi{10.48550/arXiv.2304.01874}.
\newblock URL \url{https://doi.org/10.48550/arXiv.2304.01874}.

\bibitem[Wang et~al.(2020{\natexlab{a}})Wang, Cao, Jia, and Gong]{featureflip}
Binghui Wang, Xiaoyu Cao, Jinyuan Jia, and Neil~Zhenqiang Gong.
\newblock On certifying robustness against backdoor attacks via randomized
  smoothing.
\newblock \emph{CoRR}, abs/2002.11750, 2020{\natexlab{a}}.
\newblock URL \url{https://arxiv.org/abs/2002.11750}.

\bibitem[Wang et~al.(2020{\natexlab{b}})Wang, Sreenivasan, Rajput, Vishwakarma,
  Agarwal, Sohn, Lee, and Papailiopoulos]{break_backdoorfl}
Hongyi Wang, Kartik Sreenivasan, Shashank Rajput, Harit Vishwakarma, Saurabh
  Agarwal, Jy{-}yong Sohn, Kangwook Lee, and Dimitris~S. Papailiopoulos.
\newblock Attack of the tails: Yes, you really can backdoor federated learning.
\newblock In Hugo Larochelle, Marc'Aurelio Ranzato, Raia Hadsell,
  Maria{-}Florina Balcan, and Hsuan{-}Tien Lin, editors, \emph{Advances in
  Neural Information Processing Systems 33: Annual Conference on Neural
  Information Processing Systems 2020, NeurIPS 2020, December 6-12, 2020,
  virtual}, 2020{\natexlab{b}}.
\newblock URL
  \url{https://proceedings.neurips.cc/paper/2020/hash/b8ffa41d4e492f0fad2f13e29e1762eb-Abstract.html}.

\bibitem[Wang et~al.(2021)Wang, Zhang, Xu, Lin, Jana, Hsieh, and
  Kolter]{betacrown}
Shiqi Wang, Huan Zhang, Kaidi Xu, Xue Lin, Suman Jana, Cho{-}Jui Hsieh, and
  J.~Zico Kolter.
\newblock Beta-crown: Efficient bound propagation with per-neuron split
  constraints for neural network robustness verification.
\newblock In Marc'Aurelio Ranzato, Alina Beygelzimer, Yann~N. Dauphin, Percy
  Liang, and Jennifer~Wortman Vaughan, editors, \emph{Advances in Neural
  Information Processing Systems 34: Annual Conference on Neural Information
  Processing Systems 2021, NeurIPS 2021, December 6-14, 2021, virtual}, pages
  29909--29921, 2021.
\newblock URL
  \url{https://proceedings.neurips.cc/paper/2021/hash/fac7fead96dafceaf80c1daffeae82a4-Abstract.html}.

\bibitem[Wang et~al.(2022{\natexlab{a}})Wang, Levine, and Feizi]{conjecture}
Wenxiao Wang, Alexander Levine, and Soheil Feizi.
\newblock Lethal dose conjecture on data poisoning.
\newblock \emph{CoRR}, abs/2208.03309, 2022{\natexlab{a}}.
\newblock \doi{10.48550/arXiv.2208.03309}.
\newblock URL \url{https://doi.org/10.48550/arXiv.2208.03309}.

\bibitem[Wang et~al.(2022{\natexlab{b}})Wang, Levine, and Feizi]{dpa_improve}
Wenxiao Wang, Alexander Levine, and Soheil Feizi.
\newblock Improved certified defenses against data poisoning with
  (deterministic) finite aggregation.
\newblock \emph{CoRR}, abs/2202.02628, 2022{\natexlab{b}}.
\newblock URL \url{https://arxiv.org/abs/2202.02628}.

\bibitem[Wang et~al.(2022{\natexlab{c}})Wang, Liu, Hu, Wang, Yin, and Cui]{mdr}
Xutong Wang, Chaoge Liu, Xiaohui Hu, Zhi Wang, Jie Yin, and Xiang Cui.
\newblock Make data reliable: An explanation-powered cleaning on malware
  dataset against backdoor poisoning attacks.
\newblock In \emph{Annual Computer Security Applications Conference, {ACSAC}
  2022, Austin, TX, USA, December 5-9, 2022}, pages 267--278. {ACM},
  2022{\natexlab{c}}.
\newblock \doi{10.1145/3564625.3564661}.
\newblock URL \url{https://doi.org/10.1145/3564625.3564661}.

\bibitem[Weber et~al.(2020)Weber, Xu, Karlas, Zhang, and Li]{RAB}
Maurice Weber, Xiaojun Xu, Bojan Karlas, Ce~Zhang, and Bo~Li.
\newblock {RAB:} provable robustness against backdoor attacks.
\newblock \emph{CoRR}, abs/2003.08904, 2020.
\newblock URL \url{https://arxiv.org/abs/2003.08904}.

\bibitem[Xiang and Mittal(2021)]{patchguard++}
Chong Xiang and Prateek Mittal.
\newblock Patchguard++: Efficient provable attack detection against adversarial
  patches.
\newblock \emph{CoRR}, abs/2104.12609, 2021.
\newblock URL \url{https://arxiv.org/abs/2104.12609}.

\bibitem[Xiang et~al.(2023)Xiang, Xiong, and Li]{cbd}
Zhen Xiang, Zidi Xiong, and Bo~Li.
\newblock {CBD:} {A} certified backdoor detector based on local dominant
  probability.
\newblock In Alice Oh, Tristan Naumann, Amir Globerson, Kate Saenko, Moritz
  Hardt, and Sergey Levine, editors, \emph{Advances in Neural Information
  Processing Systems 36: Annual Conference on Neural Information Processing
  Systems 2023, NeurIPS 2023, New Orleans, LA, USA, December 10 - 16, 2023},
  2023.
\newblock URL
  \url{http://papers.nips.cc/paper\_files/paper/2023/hash/0fbf046448d7eea18b982001320b9a10-Abstract-Conference.html}.

\bibitem[Xu et~al.(2020)Xu, Shi, Zhang, Wang, Chang, Huang, Kailkhura, Lin, and
  Hsieh]{lirpa}
Kaidi Xu, Zhouxing Shi, Huan Zhang, Yihan Wang, Kai{-}Wei Chang, Minlie Huang,
  Bhavya Kailkhura, Xue Lin, and Cho{-}Jui Hsieh.
\newblock Automatic perturbation analysis for scalable certified robustness and
  beyond.
\newblock In Hugo Larochelle, Marc'Aurelio Ranzato, Raia Hadsell,
  Maria{-}Florina Balcan, and Hsuan{-}Tien Lin, editors, \emph{Advances in
  Neural Information Processing Systems 33: Annual Conference on Neural
  Information Processing Systems 2020, NeurIPS 2020, December 6-12, 2020,
  virtual}, 2020.
\newblock URL
  \url{https://proceedings.neurips.cc/paper/2020/hash/0cbc5671ae26f67871cb914d81ef8fc1-Abstract.html}.

\bibitem[Yang et~al.(2023)Yang, Chi, Liu, Zhao, Huang, Cai, and
  Zhang]{shared_sat}
Pengfei Yang, Zhiming Chi, Zongxin Liu, Mengyu Zhao, Cheng{-}Chao Huang,
  Shaowei Cai, and Lijun Zhang.
\newblock Incremental satisfiability modulo theory for verification of deep
  neural networks.
\newblock \emph{CoRR}, abs/2302.06455, 2023.
\newblock \doi{10.48550/arXiv.2302.06455}.
\newblock URL \url{https://doi.org/10.48550/arXiv.2302.06455}.

\bibitem[Yatsura et~al.(2023)Yatsura, Sakmann, Hua, Hein, and Metzen]{easier}
Maksym Yatsura, Kaspar Sakmann, N.~Grace Hua, Matthias Hein, and Jan~Hendrik
  Metzen.
\newblock Certified defences against adversarial patch attacks on semantic
  segmentation.
\newblock In \emph{The Eleventh International Conference on Learning
  Representations, {ICLR} 2023, Kigali, Rwanda, May 1-5, 2023}. OpenReview.net,
  2023.
\newblock URL \url{https://openreview.net/pdf?id=b0JxQC7JLWh}.

\bibitem[Zhang et~al.(2020)Zhang, Chen, Xiao, Gowal, Stanforth, Li, Boning, and
  Hsieh]{crownibp}
Huan Zhang, Hongge Chen, Chaowei Xiao, Sven Gowal, Robert Stanforth, Bo~Li,
  Duane~S. Boning, and Cho{-}Jui Hsieh.
\newblock Towards stable and efficient training of verifiably robust neural
  networks.
\newblock In \emph{8th International Conference on Learning Representations,
  {ICLR} 2020, Addis Ababa, Ethiopia, April 26-30, 2020}. OpenReview.net, 2020.
\newblock URL \url{https://openreview.net/forum?id=Skxuk1rFwB}.

\bibitem[Zhang et~al.(2022{\natexlab{a}})Zhang, Wang, Xu, Li, Li, Jana, Hsieh,
  and Kolter]{cuttingplane}
Huan Zhang, Shiqi Wang, Kaidi Xu, Linyi Li, Bo~Li, Suman Jana, Cho{-}Jui Hsieh,
  and J.~Zico Kolter.
\newblock General cutting planes for bound-propagation-based neural network
  verification.
\newblock \emph{CoRR}, abs/2208.05740, 2022{\natexlab{a}}.
\newblock \doi{10.48550/arXiv.2208.05740}.
\newblock URL \url{https://doi.org/10.48550/arXiv.2208.05740}.

\bibitem[Zhang et~al.(2021)Zhang, Albarghouthi, and D'Antoni]{certlstm}
Yuhao Zhang, Aws Albarghouthi, and Loris D'Antoni.
\newblock Certified robustness to programmable transformations in lstms.
\newblock In Marie{-}Francine Moens, Xuanjing Huang, Lucia Specia, and
  Scott~Wen{-}tau Yih, editors, \emph{Proceedings of the 2021 Conference on
  Empirical Methods in Natural Language Processing, {EMNLP} 2021, Virtual Event
  / Punta Cana, Dominican Republic, 7-11 November, 2021}, pages 1068--1083.
  Association for Computational Linguistics, 2021.
\newblock \doi{10.18653/v1/2021.emnlp-main.82}.
\newblock URL \url{https://doi.org/10.18653/v1/2021.emnlp-main.82}.

\bibitem[Zhang et~al.(2022{\natexlab{b}})Zhang, Albarghouthi, and
  D'Antoni]{bagflip}
Yuhao Zhang, Aws Albarghouthi, and Loris D'Antoni.
\newblock Bagflip: {A} certified defense against data poisoning.
\newblock \emph{CoRR}, abs/2205.13634, 2022{\natexlab{b}}.
\newblock \doi{10.48550/arXiv.2205.13634}.
\newblock URL \url{https://doi.org/10.48550/arXiv.2205.13634}.

\bibitem[Zhu et~al.(2019)Zhu, Huang, Li, Taylor, Studer, and
  Goldstein]{triggerless_convpolytope}
Chen Zhu, W.~Ronny Huang, Hengduo Li, Gavin Taylor, Christoph Studer, and Tom
  Goldstein.
\newblock Transferable clean-label poisoning attacks on deep neural nets.
\newblock In Kamalika Chaudhuri and Ruslan Salakhutdinov, editors,
  \emph{Proceedings of the 36th International Conference on Machine Learning,
  {ICML} 2019, 9-15 June 2019, Long Beach, California, {USA}}, volume~97 of
  \emph{Proceedings of Machine Learning Research}, pages 7614--7623. {PMLR},
  2019.
\newblock URL \url{http://proceedings.mlr.press/v97/zhu19a.html}.

\end{thebibliography}
\bibliographystyle{plainnat}

%%%%%%%%%%%%%%%%%%%%%%%%%%%%%%%%%%%%%%%%%%%%%%%%%%%%%%%%%%%%%%%%%%%%%%%%%%%%%%%
%%%%%%%%%%%%%%%%%%%%%%%%%%%%%%%%%%%%%%%%%%%%%%%%%%%%%%%%%%%%%%%%%%%%%%%%%%%%%%%
% DELETE THIS PART. DO NOT PLACE CONTENT AFTER THE REFERENCES!

%%%%%%%%%%%%%%%%%%%%%%%%%%%%%%%%%%%%%%%%%%%%%%%%%%%%%%%%%%%%

\clearpage
\appendix
\section{Proof of Theorem~\ref{theorem: main}}
\label{sec: appendix_main_proof}
\begin{figure}
  \centering
  % \vspace{-1em}
  \input{figs/figure2}
  \caption{An illustration of the proof of Theorem~\ref{theorem: main}. It shows the worst case for \technique, where the attacker can change all predictions in \colorbox{gray!20}{$\dabstain$} and \colorbox{red!20}{$\dbackdoor$} to the runner-up label $\secondy$. Note that we group $\dabstain$, $\dbackdoor$, and $\dsafe$ together to ease illustration.}
  \label{fig:proof}
  % \vspace{-2em}
\end{figure}

We start by formally defining $\Nfirsty$, $\Nsecondy$, and $\Nabs$
as 
\begin{align*}
        \Nfirsty &= \sum_{i=1}^n \mathds{1}_{\modelpred{D_i}{\bfxtest} = \firsty \wedge \modelpredcert{D_i}{\bfxtest} = \certlabel}, \\
        \Nsecondy &= \sum_{i=1}^n \mathds{1}_{\modelpred{D_i}{\bfxtest} = \secondy \wedge \modelpredcert{D_i}{\bfxtest} = \certlabel}, \\
        \Nabs &= \sum_{i=1}^n \mathds{1}_{\modelpredcert{D_i}{\bfxtest} = \abslabel.}
    \end{align*}

\begin{proof}
For any poisoned dataset $\poisoned{D}$, we partition $\poisoned{D}$ into $n$ sub-datasets $\{\poisoned{D}_1, \ldots, \poisoned{D}_n\}$ according to $\{D_1, \ldots, D_n\}$ from the clean dataset $D$.
Note that we can determine such a correspondence between $D_i$ and $\poisoned{D}_i$ because our hash function is deterministic and only depends on each training example.
We further divide $\{D_1, \ldots, D_n\}$ into three disjoint parts $\dabstain$, $\dbackdoor$, and $\dsafe$ in the following way, 
\begin{itemize}
    \item $\dabstain = \{D_i \mid \modelpredcert{D_i}{\bfx} = \abslabel\}$ are the sub-datasets, on which $\alg$ abstains from making the prediction on $\bfx$. 
    From the definition of $\Nabs$, we have $|\dabstain| = \Nabs$.
    Intuitively, $\dabstain$ contains the sub-datasets that can possibly be attacked by the test input $\poisoned{\bfxtest}$ with malicious triggers.
    
    \item $\dbackdoor$ are the sub-datasets on which $\alg$ does not abstain and are also poisoned, i.e., each of them has at least one training example removed or inserted.
    Even though we do not know the exact sub-datasets in $\dbackdoor$, we know $|\dbackdoor| \le \poisonins$ because $\poisoned{D} \in \poisonset$ constrains that there are at most $\poisonins$ such poisoned sub-datasets.
    
    \item $\dsafe = \{D_i \mid D_i = \poisoned{D}_i \wedge  \modelpredcert{D_i}{\bfx} = \certlabel\}$ contains the clean sub-datasets, on which $\alg$ does not abstain. 
\end{itemize}

We denote the numbers of the original top prediction $\firsty$ and the original runner-up prediction $\secondy$ on the backdoored data $\poisoned{D}$ and $\poisoned{\bfx}$ as $\poisonedNfirsty$ and $\poisonedNsecondy$, respectively.
Formally, 
\[ \poisonedNfirsty = \sum_{i=1}^n \mathds{1}_{\modelpred{\poisoned{D}_i}{\poisoned{\bfxtest}} = \firsty},\quad
        \poisonedNsecondy = \sum_{i=1}^n \mathds{1}_{\modelpred{\poisoned{D}_i}{\poisoned{\bfxtest}} = \secondy}\]
Next, we prove Eq~\ref{eq:dab_cert} for any backdoored data $\poisoned{D}$ and $\poisoned{\bfx}$ by showing that 
\begin{align}
    \poisonedNfirsty \ge \poisonedNsecondy + \mathds{1}_{\firsty > \secondy} \label{eq: counts}
\end{align}

We prove Eq~\ref{eq: counts} by showing a lower bound of $\poisonedNfirsty$ is $\Nfirsty - \poisonins$ and an upper bound of $\poisonedNsecondy$ is $\Nsecondy + \poisonins + \Nabs$. 
Together with the definition of $r$, we can prove Eq~\ref{eq: counts} because we have,
\begin{small}
\begin{align*}
    &\poisonedNfirsty - \poisonedNsecondy - \mathds{1}_{\firsty > \secondy} \\
    \ge& \Nfirsty - \poisonins - (\Nsecondy + \poisonins + \Nabs) - \mathds{1}_{\firsty > \secondy}\\
    =&\Nfirsty - \Nsecondy -2\poisonins - \Nabs - \mathds{1}_{\firsty > \secondy}\\
    =&\Nfirsty - \Nsecondy -2\lfloor \frac{\Nfirsty - \Nsecondy - \Nabs - \mathds{1}_{\firsty > \secondy}}{2}\rfloor  - \Nabs - \mathds{1}_{\firsty > \secondy}\\
    \ge& \Nfirsty - \Nsecondy - (\Nfirsty - \Nsecondy - \Nabs - \mathds{1}_{\firsty > \secondy}) - \Nabs - \mathds{1}_{\firsty > \secondy}\\
    =& 0.
\end{align*}
\end{small}%
Note that the second last line holds iff $\Nfirsty - \Nsecondy - \Nabs - \mathds{1}_{\firsty > \secondy} \ge 0$. Otherwise, we have $\poisonins = \specialcase$.

As shown in Figure~\ref{fig:proof}, the lower bound of $\poisonedNfirsty$ can be computed by noticing that 1) the attacker can change any prediction in $\dbackdoor$ from $\firsty$ to another label because these datasets are poisoned, 2) the attacker can change any prediction in $\dabstain$ to another label because CROWN-IBP cannot certify the prediction under the evasion attacks, and 3) the attacker cannot change anything in $\dsafe$ because of the guarantee of CROWN-IBP and $\dsafe$ is not poisoned, 
\[\forall D_i \in \dsafe, \poisoned{\bfxtest} \in \perturb(\bfxtest).\ \modelpred{D_i}{\bfxtest} = \modelpred{D_i}{\poisoned{\bfxtest}} = \modelpred{\poisoned{D}_i}{\poisoned{\bfxtest}}\]
The upper bound of $\poisonedNsecondy$ can be computed by noticing that 1) the attacker can change any prediction in $\dbackdoor$ to $\secondy$, 2) the attacker can change any prediction in $\dabstain$ to $\secondy$, and 3) the attacker cannot change anything in $\dsafe$.

We complete the proof by showing that the best attack strategy of the attacker is to change the prediction of $\salg$ to the runner-up label $\secondy$.
If the attacker chooses to change the prediction of $\salg$ to another label $y''$, denoted the counts as $\poisoned{N}_{y''}$, then the upper bound of $\poisoned{N}_{y''}$ will be always smaller or equal to $\poisonedNsecondy$.
\end{proof}

\section{Proof of Theorem~\ref{theorem: inversed}}
\label{sec: proof_inversed}
\begin{proof}
    Theorem~\ref{theorem: main} tells that either $\poisonins=\specialcase$ or the following equation holds,
    \begin{align}
     \forall D' \in \poisonsetpoisoned,\ \bfxtest' \in \perturb(\poisoned{\bfxtest}).\ \smodelpred{\poisoned{D}}{\poisoned{\bfxtest}} = \smodelpred{D'}{\bfxtest'} \label{eq:dab_cert_poison}
    \end{align}

    By the symmetrical definition of $\perturbtrain$ and $\perturb$, we have 
    \begin{align}
        \forall D.\ &\poisoned{D} \in \poisonset \implies D \in \poisonsetpoisoned \label{eq: property1}\\
        \forall \bfxtest.\ &\poisoned{\bfxtest} \in \perturb(\bfxtest) \implies \bfxtest \in \perturb(\poisoned{\bfxtest}). \label{eq: property2}
    \end{align}
    Then, for all possible clean data $D$ and $\bfxtest$, we have 
    \begin{align*}
        &\poisoned{D} \in \poisonset \wedge \poisoned{\bfx} \in \perturb(\bfxtest) \\
        \implies& D \in \poisonsetpoisoned \wedge \bfxtest \in \perturb(\poisoned{\bfxtest}) \tag*{(By Eq~\ref{eq: property1} and Eq~\ref{eq: property2})}\\
        \implies& \smodelpred{\poisoned{D}}{\poisoned{\bfxtest}} = \smodelpred{D}{\bfxtest} \tag*{(By Eq~\ref{eq:dab_cert_poison})}
    \end{align*}
\end{proof}

\section{Detailed Results and Setups of Section~\ref{sec: rq1}}
\subsection{Detailed Experiment Setups}
\label{sec: appendix_detail_setup}
\paragraph{Hyper-parameters of BagFlip}
In Figure~\ref{fig:bd_cifar10_ember}, we set the bag size of BagFlip to 200, 3000, and 2000 for MNIST, EMBER, and CIFAR10 datasets, respectively.

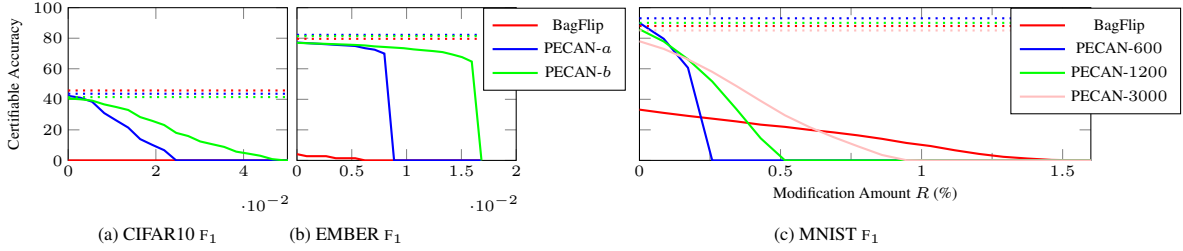
\begin{figure*}
\centering
\tiny
 \setlength{\tabcolsep}{1pt}
\begin{tabular}{cc}
    \trimbox{0cm 0cm 0.3cm 0cm}{
\pgfplotsset{filter discard warning=false}
\pgfplotsset{every axis legend/.append style={
at={(1.5,1)},
anchor=north east}} 

\pgfplotscreateplotcyclelist{whatever}{%
    red,thick,every mark/.append style={fill=blue!80!black},mark=none\\%
    blue,thick,every mark/.append style={fill=red!80!black},mark=none\\%
    green,thick,every mark/.append style={fill=blue!80!black},mark=none\\%
    red,thick,dotted,every mark/.append style={fill=blue!80!black},mark=none\\%
    blue,thick,dotted,every mark/.append style={fill=red!80!black},mark=none\\%
    green,thick,dotted,every mark/.append style={fill=blue!80!black},mark=none\\%
    % gray,thick,dashed,every mark/.append style={fill=red!80!black},mark=none\\%
    % black,thick,every mark/.append style={fill=blue!80!black},mark=none\\%
    % gray,thick,every mark/.append style={fill=red!80!black},mark=none\\%
    % % brown!60!black,every mark/.append style={fill=brown!80!black},mark=triangle*\\%
    % black,mark=star\\%
    % blue,every mark/.append style={fill=blue!80!black},mark=diamond*\\%
    % red,densely dashed,every mark/.append style={solid,fill=red!80!black},mark=*\\%
    % brown!60!black,densely dashed,every mark/.append style={solid,fill=brown!80!black},mark=square*\\%
    % black,densely dashed,every mark/.append style={solid,fill=gray},mark=triangle*\\%
    % blue,densely dashed,mark=star,every mark/.append style=solid\\%
    % red,densely dashed,every mark/.append style={solid,fill=red!80!black},mark=diamond*\\%
    }
    
\begin{tikzpicture}
    \begin{groupplot}[
            group style={
                group size=2 by 1,
                horizontal sep=.05in,
                vertical sep=.05in,
                ylabels at=edge left,
                yticklabels at=edge left,
                xlabels at=edge bottom,
                xticklabels at=edge bottom,
            },
            height=.8in,
            xlabel near ticks,
            ylabel near ticks,
            scale only axis,
            width=0.17*\textwidth,
            xmin=0,
            ymin=0,
            ymax=100,
            %xmax=1.1
        ]

        \nextgroupplot[
            % title=Robustness by demographic group,
            ylabel=Certifiable Accuracy,
            xmax=0.05,
            % xtick={0,0.02,0.04},
            % minor xtick={0.01,0.03},
            cycle list name=whatever,
            xlabel=]
        \addplot table [x=x,y=2, col sep=comma]{data/cifar10-old.csv};
        \addplot table [x=x,y=0, col sep=comma]{data/cifar10-old.csv};
        \addplot table [x=x,y=1, col sep=comma]{data/cifar10-old.csv};
        \addplot table [x=x,y=2-normal, col sep=comma]{data/cifar10-old.csv};
        \addplot table [x=x,y=0-normal, col sep=comma]{data/cifar10-old.csv};
        \addplot table [x=x,y=1-normal, col sep=comma]{data/cifar10-old.csv};
        
        \nextgroupplot[
            xmax=0.02,
            % xtick={0,0.005,0.01,0.015},
            % minor xtick={0.0025,0.0075,0.0125,0.0175},
             cycle list name=whatever,
        ]
        \addplot table [x=x,y=2, col sep=comma]{data/ember.csv};
        \addplot table [x=x,y=0, col sep=comma]{data/ember.csv};
        \addplot table [x=x,y=1, col sep=comma]{data/ember.csv};
        \addplot table [x=x,y=2-normal, col sep=comma]{data/ember.csv};
        \addplot table [x=x,y=0-normal, col sep=comma]{data/ember.csv};
        \addplot table [x=x,y=1-normal, col sep=comma]{data/ember.csv};

\legend{BagFlip, \technique-$a$, \technique-$b$};

%  \node (P1) at (pvgroup c5r1.north east) {};
%     \node (P2) at (pvgroup c5r1.south east) {};
%     \path (P1) -- node[right]{\pgfplotslegendfromname{singlelegend}} (P2);
    
    \end{groupplot}
    \node at (1.2,-1) {\scriptsize (a) CIFAR10 $\featflipone$};
    \node at (3.7,-1) {\scriptsize (b) EMBER $\featflipone$};
    % \node[draw] at (7.5,1.7) {\scriptsize (c) EMBER $\featflipone$};
    % \node[draw] at (9.8,0.5) {\scriptsize (d) EMBER $\featflipinf$};

\end{tikzpicture}
}
    & \pgfplotsset{filter discard warning=false}
\pgfplotsset{every axis legend/.append style={
at={(1.2,1)},
anchor=north east}} 

\pgfplotscreateplotcyclelist{whatever}{%
    red,thick,every mark/.append style={fill=blue!80!black},mark=none\\%
    blue,thick,every mark/.append style={fill=red!80!black},mark=none\\%
    green,thick,every mark/.append style={fill=blue!80!black},mark=none\\%
    pink,thick,every mark/.append style={fill=blue!80!black},mark=none\\%
    red,thick,dotted,every mark/.append style={fill=blue!80!black},mark=none\\%
    blue,thick,dotted,every mark/.append style={fill=red!80!black},mark=none\\%
    green,thick,dotted,every mark/.append style={fill=blue!80!black},mark=none\\%
    pink,thick,dotted,every mark/.append style={fill=blue!80!black},mark=none\\%
    % gray,thick,dashed,every mark/.append style={fill=red!80!black},mark=none\\%
    % black,thick,every mark/.append style={fill=blue!80!black},mark=none\\%
    % gray,thick,every mark/.append style={fill=red!80!black},mark=none\\%
    % % brown!60!black,every mark/.append style={fill=brown!80!black},mark=triangle*\\%
    % black,mark=star\\%
    % blue,every mark/.append style={fill=blue!80!black},mark=diamond*\\%
    % red,densely dashed,every mark/.append style={solid,fill=red!80!black},mark=*\\%
    % brown!60!black,densely dashed,every mark/.append style={solid,fill=brown!80!black},mark=square*\\%
    % black,densely dashed,every mark/.append style={solid,fill=gray},mark=triangle*\\%
    % blue,densely dashed,mark=star,every mark/.append style=solid\\%
    % red,densely dashed,every mark/.append style={solid,fill=red!80!black},mark=diamond*\\%
    }
    
\begin{tikzpicture}
    \begin{groupplot}[
            group style={
                group size=1 by 1,
                horizontal sep=.05in,
                vertical sep=.05in,
                ylabels at=edge left,
                yticklabels at=edge left,
                xlabels at=edge bottom,
                xticklabels at=edge bottom,
            },
            height=.8in,
            xlabel near ticks,
            ylabel near ticks,
            scale only axis,
            width=0.35*\textwidth,
            xmin=0,
            ymin=0,
            ymax=100,
            %xmax=1.1
            yticklabels={},
        ]

        \nextgroupplot[
            % title=Robustness by demographic group,
            ylabel=,
            xmax=1.6,
            xtick={0,0.5,1,1.5},
            minor xtick={0.25,0.75,1.25},
            cycle list name=whatever,
            xlabel=Modification Amount $\assumeradii$ (\%)]
        \addplot table [x=x,y=3, col sep=comma]{data/mnist-old.csv};
        \addplot table [x=x,y=0, col sep=comma]{data/mnist-old.csv};
        \addplot table [x=x,y=1, col sep=comma]{data/mnist-old.csv};
        \addplot table [x=x,y=2, col sep=comma]{data/mnist-old.csv};
        \addplot table [x=x,y=3-normal, col sep=comma]{data/mnist-old.csv};
        \addplot table [x=x,y=0-normal, col sep=comma]{data/mnist-old.csv};
        \addplot table [x=x,y=1-normal, col sep=comma]{data/mnist-old.csv};
        \addplot table [x=x,y=2-normal, col sep=comma]{data/mnist-old.csv};

\legend{BagFlip, \technique-$600$, \technique-$1200$, \technique-$3000$};

%  \node (P1) at (pvgroup c5r1.north east) {};
%     \node (P2) at (pvgroup c5r1.south east) {};
%     \path (P1) -- node[right]{\pgfplotslegendfromname{singlelegend}} (P2);
    
    \end{groupplot}
    % \node at (2.95,-1) {\scriptsize (a) MNIST $\featflipone$};
    % \node at (4.45,-1) {\scriptsize (b) EMBER $\featflipone$};
    % \node[draw] at (7.5,1.7) {\scriptsize (c) EMBER $\featflipone$};
    % \node[draw] at (9.8,0.5) {\scriptsize (d) EMBER $\featflipinf$};
    \node at (2.5,-1) {\scriptsize (c) MNIST $\featflipone$};

\end{tikzpicture} \\
\end{tabular}

\caption{Comparison to BagFlip on CIFAR10, EMBER, and MNIST, showing the normal accuracy (dotted lines) and the certified accuracy (solid lines) at different modification amounts $\assumeradii$. 
For CIFAR10: $a=50$ and $b=100$. For EMBER: $a=200$ and $b=400$.
}

\label{fig:bd_cifar10_ember_old}

\end{figure*}

We modified the underlying model structures of BagFlip, replacing the CNN and ResNet mentioned in their paper with fully connected neural networks to ensure a fair comparison with \technique. Figure~\ref{fig:bd_cifar10_ember_old} illustrates the comparison between \technique and BagFlip, when utilizing CNN and ResNet in BagFlip, across different datasets.
Consistent with the original paper, we maintained the bag sizes for BagFlip at 100, 3000, and 1000 for the MNIST, EMBER, and CIFAR10 datasets, respectively.

The experiment results in Figure~\ref{fig:bd_cifar10_ember_old}~(a)~and~(b) align consistently with those presented in Figure~\ref{fig:bd_cifar10_ember}~(a)~and~(b).
Figure~\ref{fig:bd_cifar10_ember_old}~(c) compares \technique and BagFlip on MNIST.
\technique achieves competitive results compared to BagFlip.
We find that the two approaches have similar normal accuracy.
Comparing \technique-$600$ and \technique-$1200$ with BagFlip,
we find that 1) \technique-$600$, \technique-$1200$, and \technique-$3000$ achieve higher certified accuracy than BagFlip when $\assumeradii \in [0,0.23]$, $\assumeradii \in [0,0.39]$, and $\assumeradii \in [0,0.63]$, respectively, and 2) BagFlip has non-zero certified accuracy when $\assumeradii \in [0.98, 1.63]$, where the certified accuracy of \technique is zero.

We argue that the gap of certified accuracy between \technique-$3000$ and BagFlip mainly comes from the different definitions of the perturbation spaces.
Moreover, the root cause of this difference is the probabilistic nature of BagFlip.

\paragraph{Training Epochs of \technique.}
As the CROWN-IBP training is sensitive to the training epochs, we report detailed training epochs we used for \technique in Table~\ref{tab:epoches}.

For BadNets experiments, we train \technique with $n=100$, 200 epochs, learning rate $5e-4$, and weight decay $5e-1$ on the CIFAR10-02 dataset and with $n=600$, 900 epochs, learning rate $5e-4$, and weight decay $5e-2$ on the MNIST dataset. 

\paragraph{Training time of \technique.} 
Like other certified approaches such as BagFlip and DPA, \technique requires training $n$ classifiers. 
For the MNIST dataset, it took \technique 5.0, 5.9, and 10.5 hours to train 600, 1200, and 3000 models, respectively, on a single V-100 GPU. 
For the CIFAR-10 dataset, \technique trained 50 and 100 models in 0.6 hours, and for the EMBER dataset, it took 1.3 and 0.6 hours to train 200 and 400 models, respectively. \technique's training is a one-time cost that can be amortized across multiple predictions.

\begin{table*}
    \centering
    \caption{Hyperparameters of \technique related to training epochs. In warm-up epochs, only the original loss function is used. In mixed-training epochs, the original loss function and CROWN-IBP loss are both used, with the weight of CROWN-IBP loss linearly increasing from 0 to 1. In full-training epochs, only the CROWN-IBP loss is used.}
    \small
    \begin{tabular}{cllll}
    \toprule
        Dataset & Epochs & Warm-up & Mixed-Training & 
Full-Training \\ \midrule
        MNIST, $l_0$, $n=600, 1200$ & 300 & 3 & 180 & 117\\
        MNIST, $l_0$, $n=3000$ & 400 & 3 & 240 & 157\\
        EMBER, $l_0$, $n=200$ & 100 & 3 & 60 & 37\\
        EMBER, $l_0$, $n=400$ & 50 & 3 & 30 & 17\\
        CIFAR10, $l_0$, $n=50, 100$ & 300 & 25 & 60 & 215\\
        $\poisoned{D}_1$ & 500 & 3 & 300 & 197\\
        $\poisoned{D}_2$ & 400 & 3 & 240 & 157\\
        $\poisoned{D}_3, \newdthree$ & 300 & 3 & 180 & 117\\
        MNIST, $l_\infty$, $n=600$ & 900 & 15 & 540 & 345\\
        MNIST, $l_\infty$, $n=1200$ & 1100 & 15 & 660 & 425\\
        CIFAR10, $l_\infty$, $n=50$ & 400 & 15 & 160 & 225\\
        CIFAR10, $l_\infty$, $n=100$ & 500 & 15 & 200 & 285\\
    \bottomrule
    \end{tabular}
    
    \label{tab:epoches}
\end{table*}

\subsection{Evaluation on the $l_\infty$ Perturbation Space}
\label{sec: appendix_exp1}
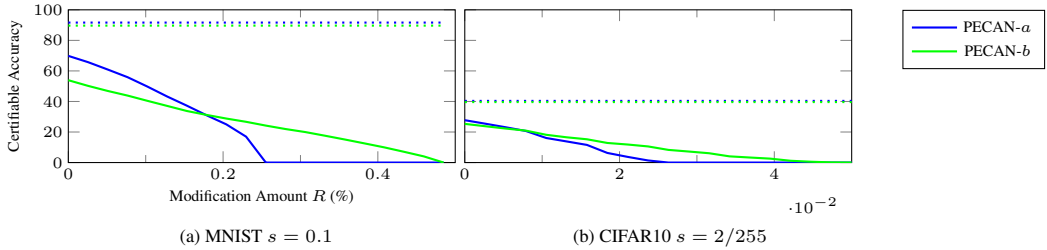
\begin{figure*}
\centering
\tiny
\pgfplotsset{filter discard warning=false}
\pgfplotsset{every axis legend/.append style={
at={(1.5,1)},
anchor=north east}} 

\pgfplotscreateplotcyclelist{whatever}{%
    blue,thick,every mark/.append style={fill=red!80!black},mark=none\\%
    green,thick,every mark/.append style={fill=blue!80!black},mark=none\\%
    blue,thick,dotted,every mark/.append style={fill=red!80!black},mark=none\\%
    green,thick,dotted,every mark/.append style={fill=blue!80!black},mark=none\\%
    red,thick,every mark/.append style={fill=blue!80!black},mark=none\\%
    red,thick,dotted,every mark/.append style={fill=blue!80!black},mark=none\\%
    pink,thick,every mark/.append style={fill=blue!80!black},mark=none\\%
    pink,thick,dotted,every mark/.append style={fill=blue!80!black},mark=none\\%
    % gray,thick,dashed,every mark/.append style={fill=red!80!black},mark=none\\%
    % black,thick,every mark/.append style={fill=blue!80!black},mark=none\\%
    % gray,thick,every mark/.append style={fill=red!80!black},mark=none\\%
    % % brown!60!black,every mark/.append style={fill=brown!80!black},mark=triangle*\\%
    % black,mark=star\\%
    % blue,every mark/.append style={fill=blue!80!black},mark=diamond*\\%
    % red,densely dashed,every mark/.append style={solid,fill=red!80!black},mark=*\\%
    % brown!60!black,densely dashed,every mark/.append style={solid,fill=brown!80!black},mark=square*\\%
    % black,densely dashed,every mark/.append style={solid,fill=gray},mark=triangle*\\%
    % blue,densely dashed,mark=star,every mark/.append style=solid\\%
    % red,densely dashed,every mark/.append style={solid,fill=red!80!black},mark=diamond*\\%
    }
    
\begin{tikzpicture}
    \begin{groupplot}[
            group style={
                group size=2 by 1,
                horizontal sep=.05in,
                vertical sep=.05in,
                ylabels at=edge left,
                yticklabels at=edge left,
                xlabels at=edge bottom,
                xticklabels at=edge bottom,
            },
            height=.8in,
            xlabel near ticks,
            ylabel near ticks,
            scale only axis,
            width=0.3*\textwidth,
            xmin=0,
            ymin=0,
            ymax=100,
            %xmax=1.1
        ]

        \nextgroupplot[
            % title=Robustness by demographic group,
            ylabel=Certifiable Accuracy,
            xmax=0.5,
            xtick={0,0.2,0.4},
            minor xtick={0.1,0.3},
            cycle list name=whatever,
            xlabel=Modification Amount $\assumeradii$ (\%)]
        % \addplot table [x=x,y=2, col sep=comma]{data/cifar10.csv};
        \addplot table [x=x,y=0, col sep=comma]{data/mnist_linf.csv};
        \addplot table [x=x,y=1, col sep=comma]{data/mnist_linf.csv};
        % \addplot table [x=x,y=2-normal, col sep=comma]{data/cifar10.csv};
        \addplot table [x=x,y=0-normal, col sep=comma]{data/mnist_linf.csv};
        \addplot table [x=x,y=1-normal, col sep=comma]{data/mnist_linf.csv};
        
        \nextgroupplot[
            xmax=0.05,
            xtick={0,0.02,0.04},
            minor xtick={0.01,0.03},
             cycle list name=whatever,
        ]
        \addplot table [x=x,y=0, col sep=comma]{data/cifar10_linf.csv};
        \addplot table [x=x,y=1, col sep=comma]{data/cifar10_linf.csv};
        \addplot table [x=x,y=0-normal, col sep=comma]{data/cifar10_linf.csv};
        \addplot table [x=x,y=1-normal, col sep=comma]{data/cifar10_linf.csv};
        % \addplot table [x=x,y=2, col sep=comma]{data/cifar10_linf.csv};
        % \addplot table [x=x,y=2-normal, col sep=comma]{data/cifar10_linf.csv};
        % \addplot table [x=x,y=3, col sep=comma]{data/cifar10_linf.csv};
        % \addplot table [x=x,y=3-normal, col sep=comma]{data/cifar10_linf.csv};

\legend{\technique-$a$, \technique-$b$}
% , ,,\technique-$a$ (SABR+$\alpha\beta$CROWN), , \technique-Cascade};

%  \node (P1) at (pvgroup c5r1.north east) {};
%     \node (P2) at (pvgroup c5r1.south east) {};
%     \path (P1) -- node[right]{\pgfplotslegendfromname{singlelegend}} (P2);
    
    \end{groupplot}
    \node at (2.5,-1) {\scriptsize (a) MNIST $\poisonfeat=0.1$};
    \node at (8,-1) {\scriptsize (b) CIFAR10 $\poisonfeat=2/255$};
    % \node[draw] at (7.5,1.7) {\scriptsize (c) EMBER $\featflipone$};
    % \node[draw] at (9.8,0.5) {\scriptsize (d) EMBER $\featflipinf$};

\end{tikzpicture}
\vspace{-1em}
\caption{Results of \technique on CIFAR10 and EMBER, showing the normal accuracy (dotted lines) and the certified accuracy (solid lines) at different modification amounts $\assumeradii$. For MNIST: $a=600$ and $b=1200$. For CIFAR10: $a=50$ and $b=100$.
}
% \vspace{-2em}
\label{fig:bd_cifar10_mnist_linf}
\end{figure*}

\paragraph{Setup} As the CROWN-IBP used in \technique can handle $\perturb$ with different $l_p$ norms, \technique can handle different $l_p$ norms as well.
We evaluate \technique on the $l_\infty$ norm with distance $\poisonfeat=0.1$ and $\poisonfeat=2/255$ on MNIST and CIFAR10, respectively, because the $l_\infty$ norm is widely applied to evaluate the robustness of image classifiers.
We use two CNN models for MNIST and CIFAR10 in this experimental setting because CROWN-IBP works better for CNN on $l_\infty$ norm than on $l_0$ norm. 
For training on MNIST and CIFAR10, we train on $\poisonfeat=0.15$ and $\poisonfeat=4/255$ but test on $\poisonfeat=0.1$ and $\poisonfeat=2/255$ to overcome the overfitting issue when $\poisonfeat$ is small, following the practice in the original paper of CROWN-IBP.

For the experiments on $l_0$ (Sections~\ref{sec: rq1}~and~\ref{sec: rq2}), we set the $\kappa_\emph{start}=0$ and $\kappa_\emph{end}=0$ for CROWN-IBP. For the experiments on $l_\infty$, we set the $\kappa_\emph{start}=1$ and $\kappa_\emph{end}=0$ for CROWN-IBP. 

\paragraph{Results}
Figure~\ref{fig:bd_cifar10_mnist_linf} shows the results of \technique against $l_\infty$ perturbation space. 
The results show that \technique achieves certified accuracy similar to $\featflipone$ as shown in Figure~\ref{fig:bd_cifar10_ember}.

\section{Detailed Results of Section~\ref{sec: rq2}}
\label{sec: appendix_rq2_detail}

\begin{table*}
\centering
\caption{Results of \technique, DPA, CROWN-IBP (C-IBP) and vanilla model without defense (NoDef) trained on three backdoored EMBER datasets. 
% We show the Wrong Prediction, Correct Prediction, and abstention rate at $\assumeradii=0.1\%$ for three partitions of the test data.
Malware with triggers is the backdoored test data that should be labeled as malware. }
\vspace{1em}
\setlength{\tabcolsep}{2.5pt}
\begin{small}
\begin{tabular}{llrrrrrrrr}
\toprule
           \multicolumn{2}{c}{Test sets}    & \multicolumn{4}{c}{Malware with triggers}                 & \multicolumn{4}{c}{Clean}\\ 
\cmidrule(lr){1-2} \cmidrule(lr){3-6} \cmidrule(lr){7-10} 
\multicolumn{2}{c}{Approaches} & \technique & DPA & C-IBP & NoDef & \technique & DPA & C-IBP & NoDef \\
\midrule
 \multirow{3}{*}{$\poisoned{D}_1$} & 
 ASR ($\downarrow$) & 3.82\% & 25.15\% & 17.01\% & 15.75\% & 4.94\% & 14.28\% & 8.47\% & 1.69\% \\
 & Accuracy ($\uparrow$)     & 49.44\% & 43.39\% & 48.27\% & 84.25\% & 72.55\% & 76.68\% & 76.86\% & 98.31\% \\
& Abstention Rate  & 46.74\% & 31.46\% & 34.72\% & N/A & 22.51\% & 9.03\% & 14.67\% & N/A  \\
\midrule
 \multirow{3}{*}{$\poisoned{D}_2$} 
& ASR ($\downarrow$) & 3.38\% & 33.87\% & 8.81\% & 44.26\% & 4.19\% & 14.33\% & 6.47\%  & 1.99\% \\
& Accuracy ($\uparrow$) & 35.27\% & 23.61\% & 40.35\% & 55.74\% & 66.36\% & 77.09\% & 70.58\% &  98.01\% \\
& Abstention Rate & 61.34\% & 42.52\% & 50.84\% & N/A & 29.45\% & 8.58\% & 22.95\% & N/A  \\
\midrule
 \multirow{3}{*}{$\poisoned{D}_3$} 
& ASR ($\downarrow$) & 2.19\% & 33.91\% & 6.64\% & 67.16\% & 1.14\% & 15.74\% & 5.69\% & 1.63\%  \\
& Accuracy ($\uparrow$) & 29.46\% & 41.89\% & 28.87\% & 32.84\% & 42.44\% & 79.06\% & 62.04\% & 98.37\%  \\
& Abstention Rate & 68.35\% & 24.20\% & 64.49\% & N/A & 56.42\% & 5.20\% & 32.27\% & N/A  \\
\bottomrule
\end{tabular}
\end{small}
\label{tab:ember_bd}
% \vspace{-1em}
\end{table*}

\begin{table}
\centering
\caption{The standard error of the mean (SEM) of CROWN-IBP and vanilla model without defense trained on three backdoored EMBER datasets. }
\setlength{\tabcolsep}{2.5pt}
\vspace{1em}
\resizebox{\columnwidth}{!}{
\small
\begin{tabular}{llrrrr}
\toprule
           \multicolumn{2}{c}{Test sets}    & \multicolumn{2}{c}{Malware with triggers}                 & \multicolumn{2}{c}{Clean}\\ 
\cmidrule(lr){1-2} \cmidrule(lr){3-4} \cmidrule(lr){5-6}
\multicolumn{2}{c}{Approaches} & C-IBP & NoDef  & C-IBP & NoDef  \\
\midrule
 \multirow{3}{*}{$\poisoned{D}_1$} & 
 SEM of ASR & 7.48 & 6.27 & 4.05 & 0.13 \\
 & SEM of Accuracy   & 8.23 & 6.27 & 1.95 & 0.13 \\
& SEM of Abs. Rate  & 6.25 & N/A & 2.78 & N/A \\
\midrule
 \multirow{3}{*}{$\poisoned{D}_2$} 
& SEM of ASR & 1.25 & 11.36 & 0.42 & 0.36 \\
& SEM of Accuracy  & 4.03 & 11.36 & 1.89 & 0.36\\
& SEM of Abs. Rate & 5.10 & N/A & 2.19 & N/A \\
\midrule
 \multirow{3}{*}{$\poisoned{D}_3$} 
& SEM of ASR & 2.25 & 11.21 & 0.50 & 0.28 \\
& SEM of Accuracy & 2.23 & 11.21 & 0.85 & 0.28 \\
& SEM of Abs. Rate & 2.29 & N/A & 1.26 & N/A \\
\bottomrule
\end{tabular}
}
\label{tab:ember_bd_std}
% \vspace{-1em}
\end{table}

To account for modifying fewer training examples than the original paper, we reduced the minimal cluster size for Isolation Forest and HDBSCAN from 0.5\% to 0.05\%. 
For MDR, we enumerated a set $\{4,5,6,7,8\}$ and set the threshold for building the graph to 7. 
For Friendly Noise, we followed the grid search for hyper-parameters (the friendly noise learning rate $\emph{lr}$ and $\mu$) in their paper~\cite{friendly_noise} and set both the noise\_eps and friendly\_clamp to 16 for CIFAR10-02, and to 8 for MNIST, and to 32 for EMBER dataset.
The best hyper-parameters sets for CIFAR10-02, MNIST, and EMBER datasets are achieved at $(\emph{lr}=10, \mu=10)$, $(\emph{lr}=100, \mu=1)$, and $(\emph{lr}=50, \mu=10)$, respectively.

For PatchGuard++ and \technique, we tune the hyper-parameter $\tau$ to the
point where the difference between accuracy and ASR is
maximized.
For the CIFAR10-02 dataset, we set $\tau=0.99$ and $\tau=0$ for \technique and PatchGuard++, respectively.
For the MNIST datsaet, we set $\tau=1$ and $\tau=0$ for \technique and PatchGuard++, respectively.
We set $\tau=0$ for PatchGuard++ because no matter what $\tau$ is, there will always be a considerable ASR.

% The standard deviation of NoDef, Isolation Forest, HDBSCAN, Spectral Signature, MDR, Friendly Noise, and CROWN-IBP can be found in Tables~\ref{tab:ember_bd_std}~and~\ref{tab:xba_res_full}. 

\subsection{Filtering Results of Empirical Defenses}
\label{sec: res_filtering}
Isolation Forest, HDBSCAN, Spectral Signature, and MDR contain filtering mechanisms that filter out poisoning examples in the training set to defend against the attack.

\begin{itemize}
    \item Isolation Forest removes \largenumber{73961} training examples, among which 146 examples are poisoned and \largenumber{73815} are false positives. \textbf{Precision: 0.20\%, Recall: 24.33\%.}
    \item HDBSCAN removes \largenumber{122323} training examples, among which 204 examples are poisoned, and \largenumber{122119} are false positives.  \textbf{Precision: 0.17\%, Recall: 34.00\%.}
    \item Spectral Signatures remove \largenumber{5000} training examples, among which 134 examples are poisoned, and \largenumber{4866} are false positives. \textbf{Precision: 2.68\%, Recall: 22.33\%.}
    \item MDR remove \largenumber{647} training examples, among which 0 examples are poisoned, and \largenumber{647} are false positives. \textbf{Precision: 0\%, Recall: 0\%.}
\end{itemize}

\subsection{Additional Comparison to DPA, CROWN-IBP, and NoDef}
\label{sec: res_non_malware}
\paragraph{Setup}
We use \citet{backdoor_Malwarepoison} to generate backdoored data by modifying 0.1\% training examples and adding triggers into the test inputs that should be labeled as malware to fool the victim model to predict the malware with malicious triggers as benign software (non-malware).  
We generate three poisoned datasets $\poisoned{D}_1, \poisoned{D}_2, \poisoned{D}_3$ and their corresponding test sets with triggers by perturbations $\featflipone$, $\featfliptwo$, and $\featflipthree$, which allow the attacker to modify up to one, two, and three features in an example, respectively. 

% \begin{figure*}[t]
%   \centering
%     \begin{tabular}{cc}
%        \input{figs/malware_res_chart.tex}  &  \input{figs/malware_res_chart1.tex}
%     \end{tabular}
%   \caption{Results of \technique, DPA, CROWN-IBP (C-IBP), and vanilla model without defense (NoDef) trained on three poisoned EMBER datasets when evaluated on (left) the malware test set with malicious triggers and (right) the (original) malware test set without malicious triggers. We note that NoDef does not have abstention rates because it does not use any defense.}
%   \label{fig: malware_res_chart}
%   % \vspace{-1em}
% \end{figure*}

% \begin{figure}[t]
%     \input{figs/malware_res_chart2.tex}
%     \caption{Results of \technique, DPA, CROWN-IBP (C-IBP), and vanilla model without defense (NoDef) trained on three poisoned EMBER datasets when evaluated on the (original) non-malware test set without triggers.}
%     \label{fig: malware_res_chart3}
% \end{figure}

Table~\ref{tab:ember_bd} show that across different trigger sizes, \textbf{\technique can defend against the backdoor attack, but DPA and CROWN-IBP fail to defend against the backdoor attack.}

% Table~\ref{tab:ember_bd} show that NoDef has the highest accuracy of \todored{XXX} on the clean test set without trigger than all three defenses.
% \technique achieves a lower ASR of $5.54\%$ than DPA ($7.85\%$), but higher than CROWN-IBP ($4.89\%$). 

\subsection{Detailed Comparison to DPA}
\label{sec: appendix_exp2}
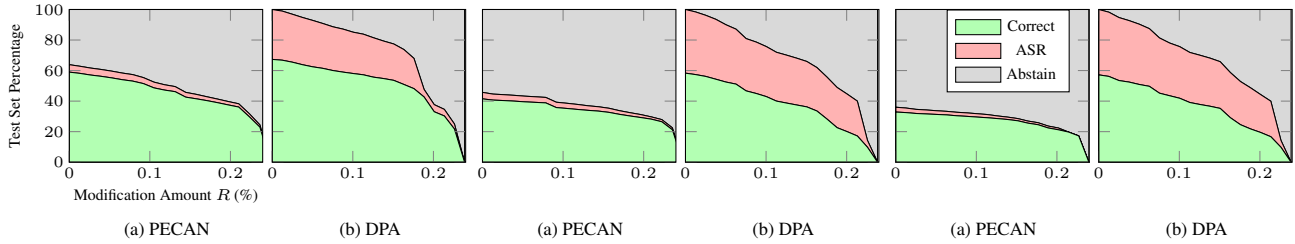
\begin{figure*}[t]
\setlength{\tabcolsep}{1pt}
    \begin{tabular}{ccc}
         \trimbox{0.3cm 0.4cm 0cm 0cm}{ 
\centering
\tiny
\pgfplotsset{filter discard warning=false}
\pgfplotsset{every axis legend/.append style={
at={(1.4,1)},
anchor=north east}} 

\pgfplotsset{
/pgfplots/area cycle list/.style={/pgfplots/cycle list={%
{black,fill=green!30!white,mark=none},%
{black,fill=red!30!white,mark=none},%
{black,fill=gray!30!white,mark=none},%
}
},
}
    
\begin{tikzpicture}
    \begin{groupplot}[
            group style={
                group size=2 by 1,
                horizontal sep=.05in,
                vertical sep=.05in,
                ylabels at=edge left,
                yticklabels at=edge left,
                xlabels at=edge bottom,
                xticklabels at=edge bottom,
            },
            height=.8in,
            xlabel near ticks,
            ylabel near ticks,
            scale only axis,
            width=0.15*\textwidth,
            xmin=0,
            ymin=0,
            ymax=100,
            stack plots=y,%   
            area style,
            %xmax=1.1
        ]

        \nextgroupplot[
            % title=Robustness by demographic group,
            ylabel=Test Set Percentage,
            xmax=0.24,
            % xtick={0,0.04,0.08},
            % minor xtick={0.02,0.06,0.1},
            xlabel=Modification Amount $\assumeradii$ (\%)]
        \addplot table [mark=none,x=x,y=y1, col sep=comma]{data/DBA_1_positive_2.csv}\closedcycle;
        \addplot table [mark=none,x=x,y=y2, col sep=comma]{data/DBA_1_positive_2.csv}\closedcycle;
        \addplot table [mark=none,x=x,y=y3, col sep=comma]{data/DBA_1_positive_2.csv}\closedcycle;
        
        \nextgroupplot[
            xmax=0.24,
            % xtick={0,0.04,0.08},
            % minor xtick={0.02,0.06,0.1},
        ]
        \addplot table [mark=none,x=x,y=y1, col sep=comma]{data/DPA_1_positive_2.csv}\closedcycle;
        \addplot table [mark=none,x=x,y=y2, col sep=comma]{data/DPA_1_positive_2.csv}\closedcycle;
        \addplot table [mark=none,x=x,y=y3, col sep=comma]{data/DPA_1_positive_2.csv}\closedcycle;

% \legend{Correct, ASR, Abstain};

%  \node (P1) at (pvgroup c5r1.north east) {};
%     \node (P2) at (pvgroup c5r1.south east) {};
%     \path (P1) -- node[right]{\pgfplotslegendfromname{singlelegend}} (P2);
    
    \end{groupplot}
    \node at (1.3,-0.9) {\scriptsize (a) \technique};
    \node at (4,-0.9) {\scriptsize (b) DPA};
    % \node[draw] at (7.5,1.7) {\scriptsize (c) EMBER $\featflipone$};
    % \node[draw] at (9.8,0.5) {\scriptsize (d) EMBER $\featflipinf$};

\end{tikzpicture}
% \begin{tikzpicture}[scale=1]
% 	\tikzstyle{every node}=[font=\scriptsize]
 
% % % 	\draw[help lines,step=5mm,gray!20] (-0.5,0) grid (7,-4);
% %     %%%%%%%%%%%%%%%%%%%%%% Inputs %%%%%%%%%%%%%%%%%%%%%%
% %     \node[inner sep=0pt] (fig) at (0,0) {\includegraphics[scale=0.305]{figs/backdoor_malware_3.pdf}};
% %     \node at (-1.7,-1.75) {\scriptsize (a) \technique};
% %     \node at (2,-1.75) {\scriptsize (b) DPA};
% \end{tikzpicture}
}& \trimbox{0.3cm 0.4cm 0cm 0cm}{ 
\centering
\tiny
\pgfplotsset{filter discard warning=false}
\pgfplotsset{every axis legend/.append style={
at={(1.4,1)},
anchor=north east}} 

\pgfplotsset{
/pgfplots/area cycle list/.style={/pgfplots/cycle list={%
{black,fill=green!30!white,mark=none},%
{black,fill=red!30!white,mark=none},%
{black,fill=gray!30!white,mark=none},%
}
},
}
    
\begin{tikzpicture}
    \begin{groupplot}[
            group style={
                group size=2 by 1,
                horizontal sep=.05in,
                vertical sep=.05in,
                ylabels at=edge left,
                yticklabels at=edge left,
                xlabels at=edge bottom,
                xticklabels at=edge bottom,
            },
            height=.8in,
            xlabel near ticks,
            ylabel near ticks,
            scale only axis,
            width=0.15*\textwidth,
            xmin=0,
            ymin=0,
            ymax=100,
            stack plots=y,%   
            area style,
            %xmax=1.1
            yticklabels={},
        ]

        \nextgroupplot[
            % title=Robustness by demographic group,
            % ylabel=Test Set Percentage,
            xmax=0.24,
            % xtick={0,0.04,0.08},
            % minor xtick={0.02,0.06,0.1},
            % xlabel=Modification Amount $\assumeradii$ (\%)
            ]
        \addplot table [mark=none,x=x,y=y1, col sep=comma]{data/DBA_2_positive_2.csv}\closedcycle;
        \addplot table [mark=none,x=x,y=y2, col sep=comma]{data/DBA_2_positive_2.csv}\closedcycle;
        \addplot table [mark=none,x=x,y=y3, col sep=comma]{data/DBA_2_positive_2.csv}\closedcycle;
        
        \nextgroupplot[
            xmax=0.24,
            % xtick={0,0.04,0.08},
            % minor xtick={0.02,0.06,0.1},
        ]
        \addplot table [mark=none,x=x,y=y1, col sep=comma]{data/DPA_2_positive_2.csv}\closedcycle;
        \addplot table [mark=none,x=x,y=y2, col sep=comma]{data/DPA_2_positive_2.csv}\closedcycle;
        \addplot table [mark=none,x=x,y=y3, col sep=comma]{data/DPA_2_positive_2.csv}\closedcycle;

% \legend{Correct, ASR, Abstain};

%  \node (P1) at (pvgroup c5r1.north east) {};
%     \node (P2) at (pvgroup c5r1.south east) {};
%     \path (P1) -- node[right]{\pgfplotslegendfromname{singlelegend}} (P2);
    
    \end{groupplot}
    \node at (1.3,-0.9) {\scriptsize (a) \technique};
    \node at (4,-0.9) {\scriptsize (b) DPA};
    % \node[draw] at (7.5,1.7) {\scriptsize (c) EMBER $\featflipone$};
    % \node[draw] at (9.8,0.5) {\scriptsize (d) EMBER $\featflipinf$};

\end{tikzpicture}
% \begin{tikzpicture}[scale=1]
% 	\tikzstyle{every node}=[font=\scriptsize]
 
% % % 	\draw[help lines,step=5mm,gray!20] (-0.5,0) grid (7,-4);
% %     %%%%%%%%%%%%%%%%%%%%%% Inputs %%%%%%%%%%%%%%%%%%%%%%
% %     \node[inner sep=0pt] (fig) at (0,0) {\includegraphics[scale=0.305]{figs/backdoor_malware_3.pdf}};
% %     \node at (-1.7,-1.75) {\scriptsize (a) \technique};
% %     \node at (2,-1.75) {\scriptsize (b) DPA};
% \end{tikzpicture}
}& \trimbox{0.3cm 0.4cm 0cm 0cm}{ 
\centering
\tiny
\pgfplotsset{filter discard warning=false}
\pgfplotsset{every axis legend/.append style={
at={(-0.15,1)},
anchor=north east}} 

\pgfplotsset{
/pgfplots/area cycle list/.style={/pgfplots/cycle list={%
{black,fill=green!30!white,mark=none},%
{black,fill=red!30!white,mark=none},%
{black,fill=gray!30!white,mark=none},%
}
},
}
    
\begin{tikzpicture}
    \begin{groupplot}[
            group style={
                group size=2 by 1,
                horizontal sep=.05in,
                vertical sep=.05in,
                ylabels at=edge left,
                yticklabels at=edge left,
                xlabels at=edge bottom,
                xticklabels at=edge bottom,
            },
            height=.8in,
            xlabel near ticks,
            ylabel near ticks,
            scale only axis,
            width=0.15*\textwidth,
            xmin=0,
            ymin=0,
            ymax=100,
            stack plots=y,%   
            area style,
            %xmax=1.1
            yticklabels={},
        ]

        \nextgroupplot[
            % title=Robustness by demographic group,
            % ylabel=Test Set Percentage,
            xmax=0.24,
            % xtick={0,0.04,0.08},
            % minor xtick={0.02,0.06,0.1},
            % xlabel=Modification Amount $\assumeradii$ (\%)
            ]
        \addplot table [mark=none,x=x,y=y1, col sep=comma]{data/DBA_3_positive_2.csv}\closedcycle;
        \addplot table [mark=none,x=x,y=y2, col sep=comma]{data/DBA_3_positive_2.csv}\closedcycle;
        \addplot table [mark=none,x=x,y=y3, col sep=comma]{data/DBA_3_positive_2.csv}\closedcycle;
        
        \nextgroupplot[
            xmax=0.24,
            % xtick={0,0.04,0.08},
            % minor xtick={0.02,0.06,0.1},
        ]
        \addplot table [mark=none,x=x,y=y1, col sep=comma]{data/DPA_3_positive_2.csv}\closedcycle;
        \addplot table [mark=none,x=x,y=y2, col sep=comma]{data/DPA_3_positive_2.csv}\closedcycle;
        \addplot table [mark=none,x=x,y=y3, col sep=comma]{data/DPA_3_positive_2.csv}\closedcycle;

\legend{Correct, ASR, Abstain};

%  \node (P1) at (pvgroup c5r1.north east) {};
%     \node (P2) at (pvgroup c5r1.south east) {};
%     \path (P1) -- node[right]{\pgfplotslegendfromname{singlelegend}} (P2);
    
    \end{groupplot}
    \node at (1.3,-0.9) {\scriptsize (a) \technique};
    \node at (4,-0.9) {\scriptsize (b) DPA};
    % \node[draw] at (7.5,1.7) {\scriptsize (c) EMBER $\featflipone$};
    % \node[draw] at (9.8,0.5) {\scriptsize (d) EMBER $\featflipinf$};

\end{tikzpicture}
% \begin{tikzpicture}[scale=1]
% 	\tikzstyle{every node}=[font=\scriptsize]
 
% % % 	\draw[help lines,step=5mm,gray!20] (-0.5,0) grid (7,-4);
% %     %%%%%%%%%%%%%%%%%%%%%% Inputs %%%%%%%%%%%%%%%%%%%%%%
% %     \node[inner sep=0pt] (fig) at (0,0) {\includegraphics[scale=0.305]{figs/backdoor_malware_3.pdf}};
% %     \node at (-1.7,-1.75) {\scriptsize (a) \technique};
% %     \node at (2,-1.75) {\scriptsize (b) DPA};
% \end{tikzpicture}
}\\
    \end{tabular}
    
    \caption{Comparison between \technique and DPA trained on $\poisoned{D}_1$, $\poisoned{D}_2$, and $\poisoned{D}_3$ across all modification amount $\assumeradii$ when evaluated on the malware test set with triggers.}
    \label{fig: dpa_compare}
\end{figure*}

Figure~\ref{fig: dpa_compare} shows the comparison between \technique and DPA on $\poisoned{D}_1$, $\poisoned{D}_2$ and $\poisoned{D}_3$.
We can observe that PECAN has much higher ASRs than DPA across all modification amounts on $\poisoned{D}_1$, $\poisoned{D}_2$ and $\poisoned{D}_3$.

\subsection{Comparison to FPA under the XBA on the EMBER dataset}
\label{sec: appendix_exp3}
\begin{table*}
    \centering
    \vspace{-1em}
    \caption{Results of \technique and FPA trained on poisoned datasets $\poisoned{D}_3$ and $\newdthree$ when evaluated on the malware test set with malicious triggers.}
    \vspace{1em}
    \resizebox{0.6\textwidth}{!}{
    \setlength{\tabcolsep}{2pt}
    \scriptsize
    \begin{tabular}{lllllll}
    \toprule
       & \multicolumn{3}{c}{$\poisoned{D}_3$}                 & \multicolumn{3}{c}{$\newdthree$}\\ 
 \cmidrule(lr){2-4} \cmidrule(lr){5-7}
        Approaches & ASR ($\downarrow$) & Accuracy ($\uparrow$) & Abs. Rate & ASR ($\downarrow$) & Accuracy ($\uparrow$) & Abs. Rate  \\ \midrule
        FPA & 
        0.72\% & 34.28\% & 65.00\% &
0.28\% & 13.07\% & 86.65\% \\ 
        \technique  &
        2.19\% & 29.46\% & 68.35\% 
        & 1.51\% & 36.74\% & 61.75\% \\ 
        \bottomrule
    \end{tabular}
    }
    \label{tab:xba_res_fpa}
    \vspace{-1em}
\end{table*}
% \paragraph{Comparison to FPA} FPA is a certified defense against backdoor attacks. 
% Compared to \technique, FPA has the following limitations.
% First, FPA cannot defend against data poisoning attacks that involve modifications to training labels or the insertion/removal of training examples.
% In contrast, \technique does not have this limitation because its perturbation space of the dataset captures such data poisoning attacks.
% Second, it is challenging for FPA to defend against dynamic backdoor attacks~\cite{dynamic_trigger} that can place fixed-size dynamic triggers on any features. 
% \technique does not have this limitation because its perturbation space of the test input effectively captures the attack space of dynamic backdoor attacks. 

The backdoor attack we used, XBA, is a \emph{clean-label} attack and only \emph{modifies} the data by placing a \emph{fixed} trigger on training and test inputs.
Thus, FPA is more suitable to defend against this attack than \technique is in principle.
To make a fair comparison between FPA and \technique, we extend XBA as a dynamic backdoor attack, placing three different triggers with size three on inputs and resulting in a poisoned dataset $\newdthree$.
To defend $\poisoned{D}_3$ and $\newdthree$, we set the number of partitions in FPA to 8 and 23 and present its results for the trigger size 3 and $9=3 \times 3$, respectively.

Table~\ref{tab:xba_res_fpa} compares FPA and \technique. 
As certified approaches, both effectively defend against backdoor attacks and achieve a low ASR on two datasets.
FPA has a higher accuracy than \technique on $\poisoned{D}_3$ because FPA is more effective under a fixed trigger.
However, on $\newdthree$, FPA has a lower accuracy than \technique, because FPA is less effective when there are different triggers in the poisoned data.

\end{document}